\documentclass{elsart}
\usepackage{graphicx}
\usepackage{subfigure}
  \def\goodgap{%
    \hspace{\subfigtopskip}%
    \hspace{\subfigbottomskip}}
\usepackage[nomarkers]{endfloat}
\begin{document}
    \runauthor{K. Korcsak-Gorzo \textit{et. al.}}
    \begin{frontmatter}
        \title{The Optical Alignment System of the ZEUS 
MicroVertex Detector}
        \author{K. Korcsak-Gorzo,}
        \author{G. Grzelak\thanksref{wars},}
        \author{K. Oliver,}
        \author{M. Dawson,}
        \author{R. Devenish\corauthref{cor},}
        \author{J. Ferrando,}
        \author{T. Matsushita\thanksref{ic},}
        \author{P. Shield\thanksref{ret},}
        \author{R. Walczak}
\thanks[wars]{Now at the University of Warsaw.}
\thanks[ic]{Now at Imperial College, London.}
\thanks[ret]{Retired.}
\corauth[cor]{Corresponding author. Email: r.devenish@physics.ox.ac.uk}        
        \address{Department of Physics, University of Oxford,\\
          Denys Wilkinson Building, Keble Road, Oxford OX1 3RH}
       \date{May 2007}
\begin{abstract}
The laser alignment system of the ZEUS microvertex detector 
is described. The detector was installed in 2001 as part of an upgrade
programme in preparation for the second phase of electron-proton physics
at the HERA collider. The alignment system monitors the position of the 
vertex detector support structure with respect to the central tracking detector 
using semi-transparent amorphous-silicon sensors and diode lasers. 
The system is fully integrated into the 
general environmental monitoring of the ZEUS detector and data has been
collected over a period of 5 years. The primary aim of defining periods of
stability for track-based alignment has been achieved and the system is able
to measure movements of the support structure to a precision around $10\,\mu$m.
\end{abstract}
\end{frontmatter}
        
%\input intro-mvd
%%
%% introduction
%%
\section{Introduction}

A silicon-strip MicroVertex Detector (MVD) was added to the ZEUS
detector in 2001 as part of an upgrade programme for high luminosity
running with the HERA-II electron-proton collider at DESY\cite{upgrade}. 
One of the main physics motivations was to improve the study of
heavy flavour production at HERA, particularly charm. With mean decay lengths 
of the order of $100\,\mu$m for charmed hadrons, precise alignment of the MVD 
active elements is crucial if their intrinsic spatial resolution of $10\,\mu$m 
and better is to be properly exploited. Alignment has been addressed in three
stages: i) during construction the position of the silicon strip detectors was
measured with respect to the local support structure using an accurate 3-D 
measuring machine; ii) an optical alignment system tracks large movements of the
MVD support structure; iii) individual MVD sensors are aligned precisely using 
charged-particle tracks from HERA run data. This paper describes the 
laser alignment system used for the second stage and summarises its performance. 
The primary aims of the laser system are to track large movements 
(at the level of $100\,\mu$m) of the MVD support structure with respect to the
central tracking detector (CTD) and to define periods 
of stability for the track alignment. A prototype \cite{mvdla1} of the system 
described here was tested and used during the construction of the MVD. 

The paper is organised as follows: the MVD is described very briefly in the 
next section; the optical alignment system is described in 
section~\ref{sec:hardware}; the readout and online control system is described in 
section~\ref{sec:readout}; data reduction and reconstruction is covered in
section~\ref{sec:recons}; the results are summarised in section~\ref{sec:results}
and the paper concludes with a short summary (section~\ref{sec:summary}).

\section{The ZEUS Microvertex Detector}
\label{sec:mvd}

\begin{figure}
    \begin{center}
        \includegraphics[width=\textwidth]{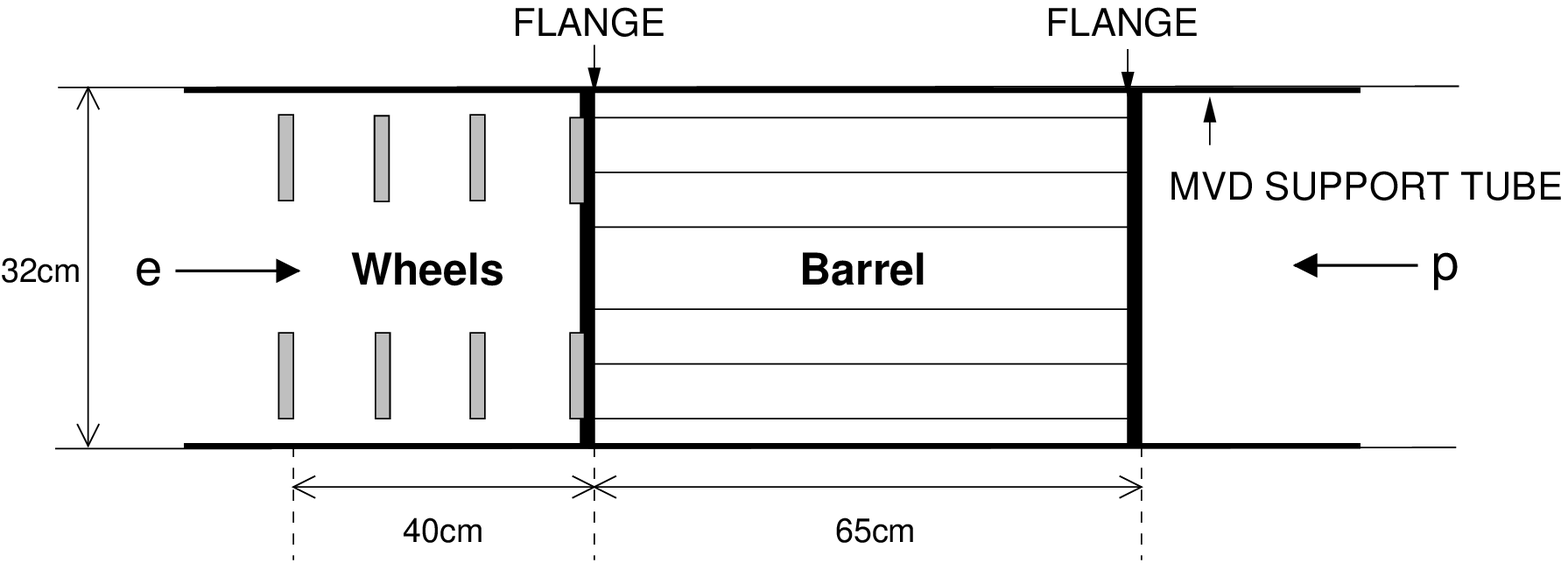}
        \caption{Layout of the ZEUS MVD along beam axis.  The barrel part
        covers the interaction region.  Four wheels cover the forward direction of
        the proton beam.} 
        \label{fig:mvdlayout}
    \end{center}
\end{figure}

\begin{figure}
    \begin{center}
        \includegraphics[scale=0.37]{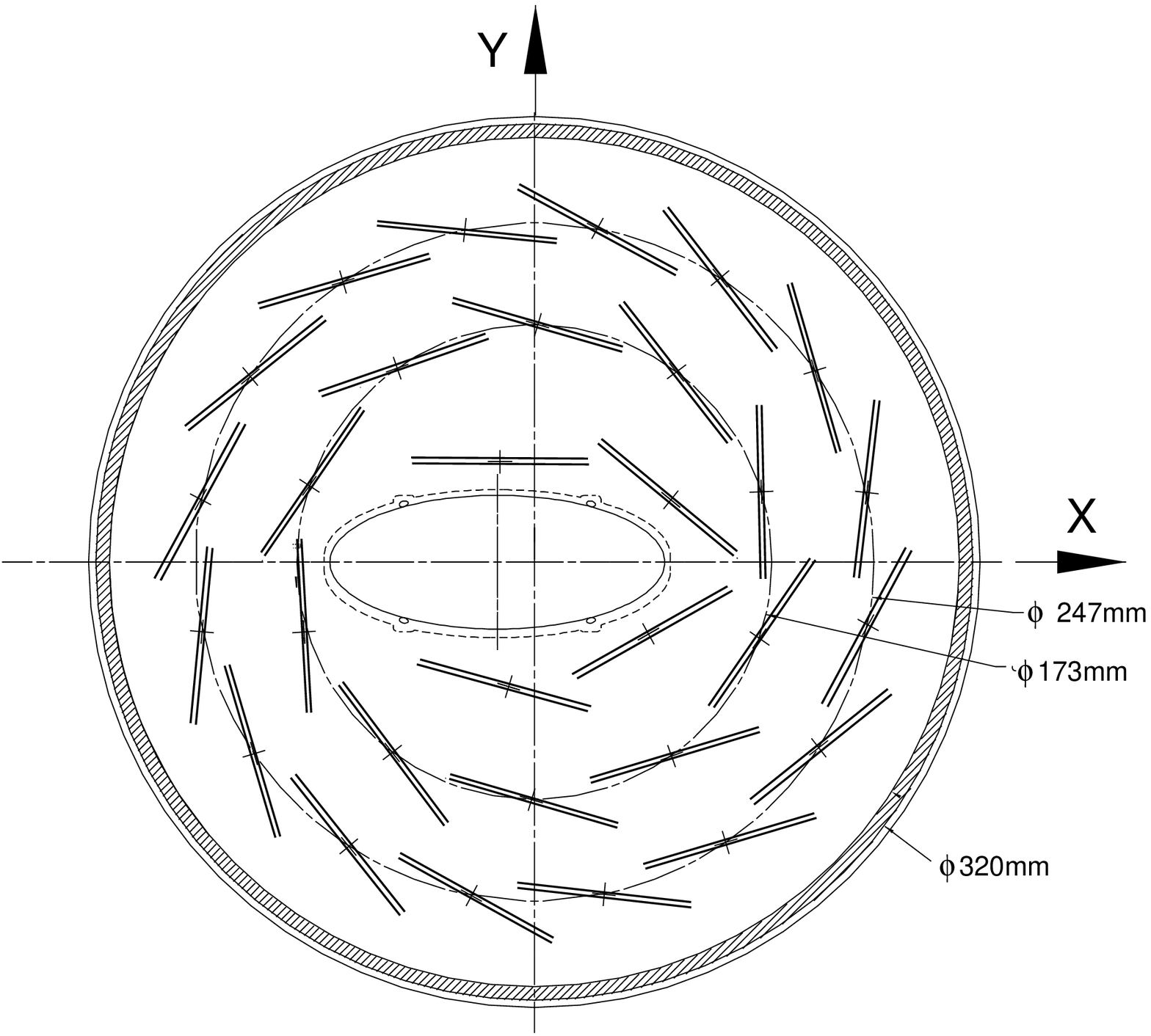}
        \goodgap
        \includegraphics[scale=0.34]{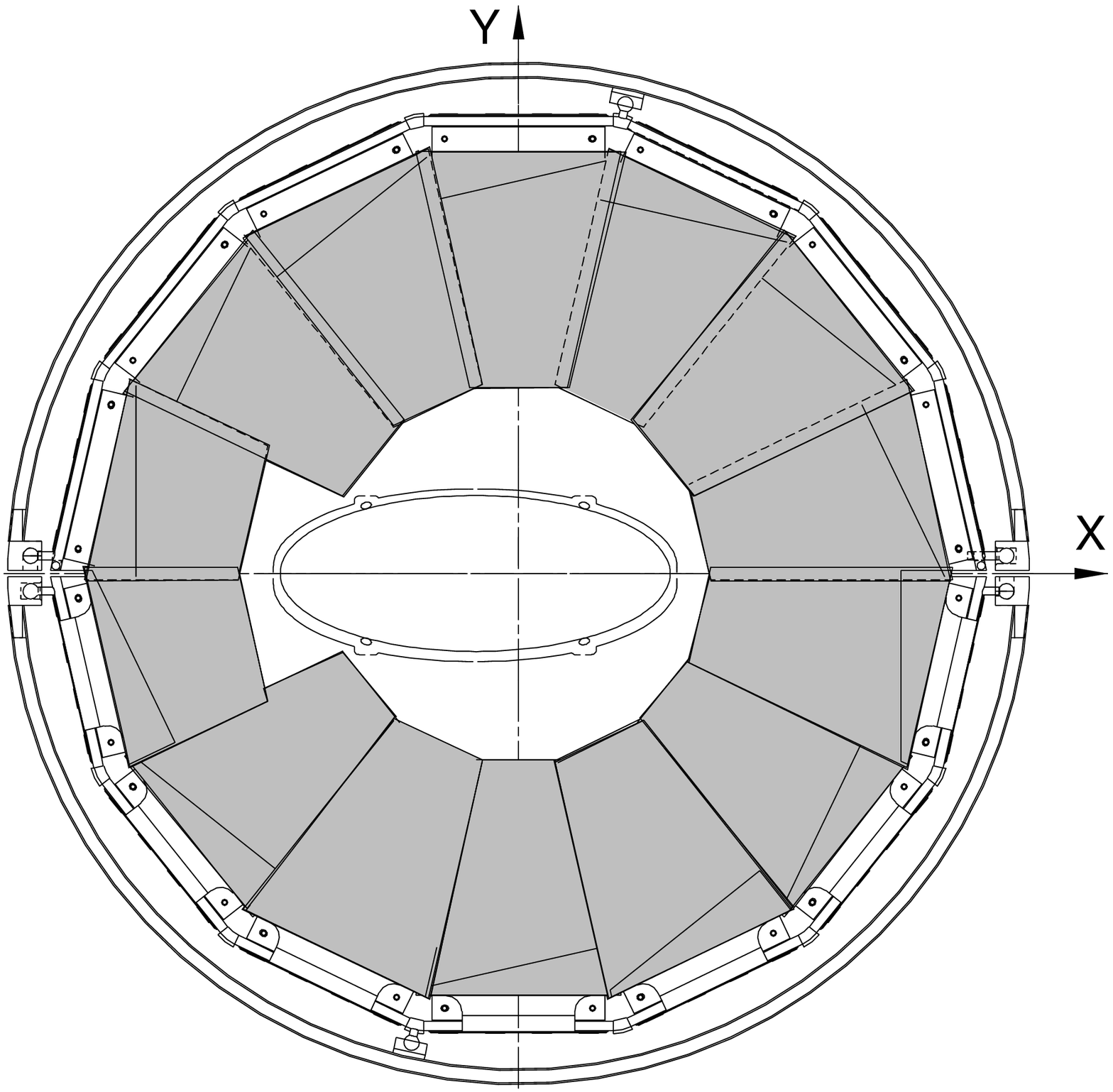}
        \caption{Cross sections of the MVD. Left and right are barrel and
        wheel, respectively. The coordinate system is that of the
ZEUS experiment, with the $z$-axis along the proton beam direction, the $x$-axis
pointing towards the centre of the HERA ring and the $y$-axis vertical.}
        \label{fig:mvdxsec}
    \end{center}
\end{figure}

The space available for the MVD is limited by the CTD
and the shape of the beam pipe.  The space inside the tracking
detector has a length of about $2\,$m and a diameter of $32\,$cm. The design
requirements of the MVD were: polar angular coverage of $10^\circ - 170^\circ$;
at least three measurements, in two projections, per track; at least $20\,\mu$m
hit resolution and impact parameter resolution of around $100\,\mu$m at $90^\circ$ for
tracks with momentum of at least $2\,$GeV/c. In order to meet these
requirements within the limited space, the MVD consists of two
parts, barrel and wheels, which are supported by a carbon-fibre tube made in two 
half-cylinders. The barrel has three concentric layers of silicon sensors but only the 
forward region is instrumented with sensors mounted on four wheels, numbered from
0 to 3.
This follows from the unequal HERA beam energies, with 27.5~GeV electrons and 
920~GeV protons, reaction products are boosted along the forward
proton direction. The layout of the MVD is shown in Fig.~\ref{fig:mvdlayout} and
Fig.~\ref{fig:mvdxsec} shows cross sections of the barrel and a wheel. 
All the MVD services and readout connections are made through the
rear end of the detector. The region at the rear of the MVD (shown to the right 
of the barrel in Fig.~\ref{fig:mvdlayout}) is used for cooling water distribution 
manifolds. 
The MVD is described in more detail in references \cite{zeusmvd1}
and \cite{zeusmvd2}. As the MVD has to fit into
an existing detector, getting services in and out is not easy. The route that
cables follow is close to the rear beampipe, through the beam hole in the
rear tracking detector and rear calorimeter. A further challenge is that, as part
of the measures to increase the luminosity for HERA-II, a superconducting 
combined-function magnet (the HERA GG magnet) penetrates the detector 
around the rear beampipe. Getting all the MVD cables to fit between this magnet
and the rear calorimeter was particularly challenging. The special very thin cables
designed to keep material at a minimum within the detector run for about five metres
from the MVD to four cable patch-boxes located above the first HERA magnets outside 
the ZEUS detector on the rear (upstream proton) side. From here much more robust cables 
take services and signals to the MVD services and readout racks about $20\,$m
away.  

%\input hardware
%%
%% optical alignment system
%%
\section{Optical Alignment System}\label{sec:hardware}

\begin{figure}
    \begin{center}
        \includegraphics[width=\textwidth]{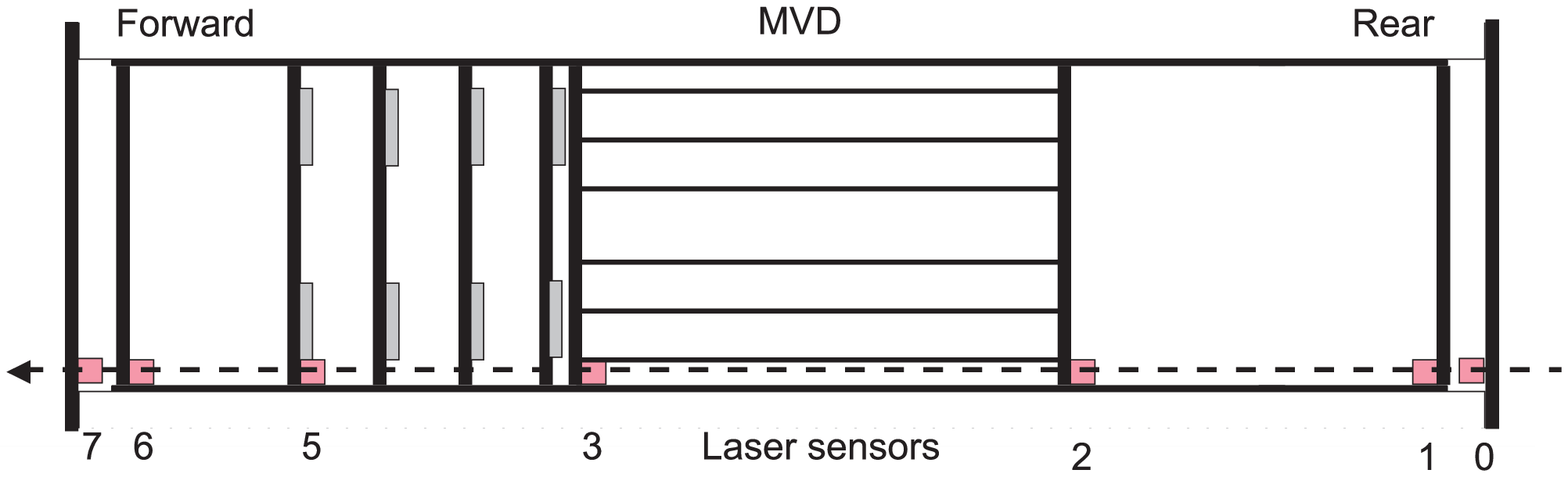}
        \caption{Schematic diagram of one laser alignment beam and sensors. Forward 
and Rear refer to the orientation of the tracking detectors, with forward in the
direction of the HERA proton beam. The sensor numbering is also shown.}
        \label{fig:la-scheme}
    \end{center}
\end{figure}

The MVD laser alignment system consists of five straightness monitors placed 
around the circumference of the MVD support tube. Each straightness monitor consists
of a collimated laser beam, approximately parallel to the collider beam line, 
with seven semi-transparent silicon sensors positioned along its path.   
Two of the sensors, mounted on the forward and rear CTD end-plates, define the 
line for that laser beam. The five remaining sensors are mounted within the MVD 
at the forward and rear end-flanges, the forward and rear barrel-flanges and the 
support structure for wheel-3, as shown schematically in Fig.~\ref{fig:la-scheme}.
Inside the MVD a laser beam is contained between sensors within a narrow carbon-fibre 
cover with a semicircular profile, glued to the inner surface of the outer 
half-cylinders. 
Each sensor provides two mutually orthogonal position measurements, with resolution
better than $10\,\mu$m, in a local sensor coordinate system. 
The alignment system is sensitive to rotations, twists and sags of the MVD support 
structure with respect to the CTD from which it is mounted. The system is not 
sensitive to translations along the beam direction.

\subsection{Position sensors and signal cables}\label{ssec:sens-cab}

The laser position sensors, DPSD-516 transparent silicon diodes (TSD), 
use semi-transparent amorphous-silicon 
as the active material. The sensors were developed by H.~Kroha 
\textit{et.  al.}\ \cite{kroha} and manufactured by EG\&G Heimann 
Optoelectronics.  The active material of the TSD sensors has a
thickness of $\sim 1\,\mu$m and an area of $5\times 5\,$mm$^2$. Signals are
read out by strips made of $\sim100\,$nm thick indium-tin-oxide, with
16 strips on each side of the amorphous-silicon. The strips on opposite
sides are perpendicular to each other, with strip pitch $312\,\mu$m and strip
gap $10\,\mu$m.  The whole structure is deposited on a $0.5\,$mm thick glass
substrate.  The sensor is transparent to light with wavelength greater than 
$\sim600\,$nm. Transmission reaches a rough plateau of 80\% at wavelengths around
$700\,$nm and above, as shown in Fig. 6 of reference \cite{mvdla1}. However
the sensitivity of the sensor drops fairly rapidly above $700\,$nm. Full details of the 
sensor and its performance are given in reference \cite{kroha}. 

\begin{figure}
    \begin{center}
      \begin{tabular}{c}
        \includegraphics[width=\textwidth,angle=270.,scale=0.5]{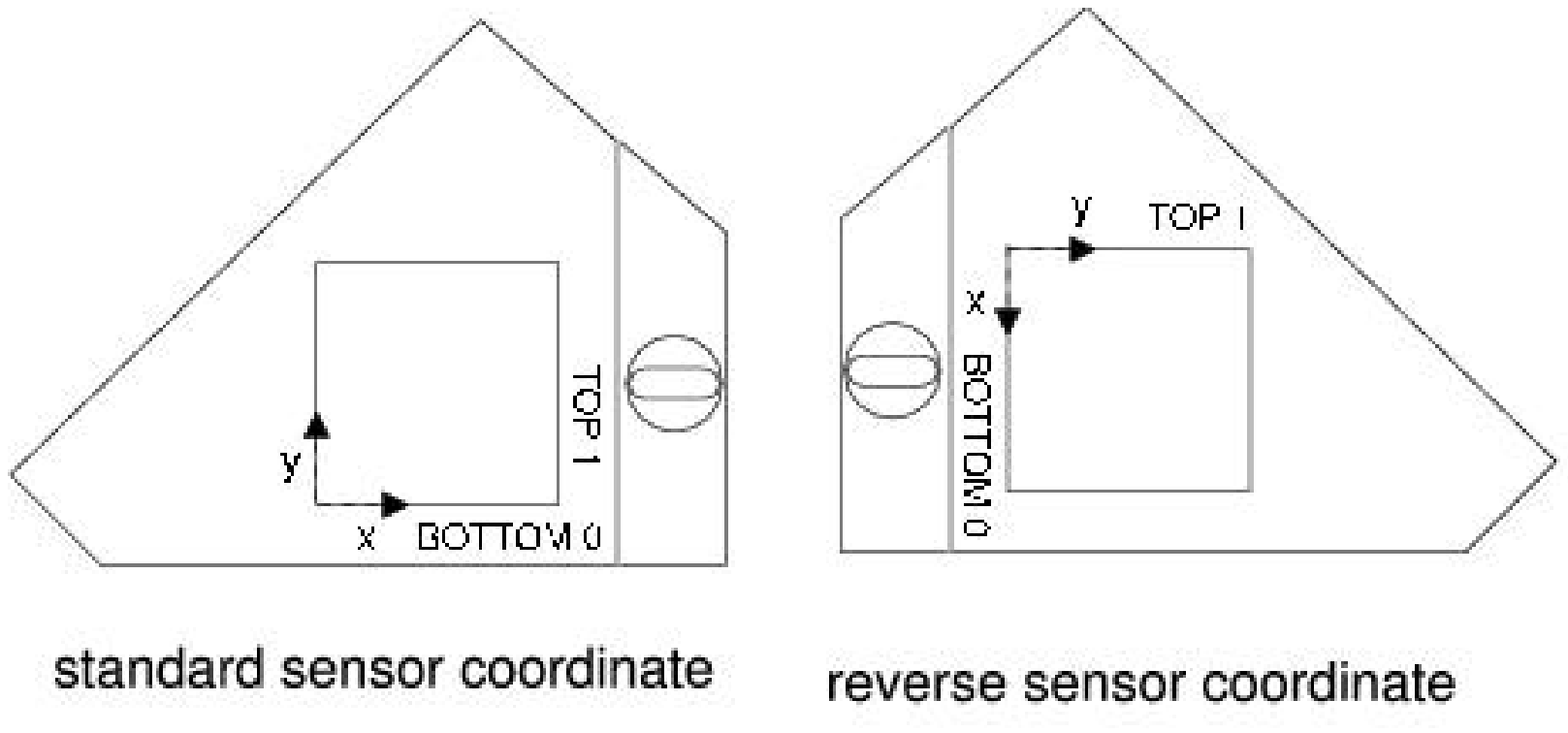} \\
        \includegraphics[width=0.70\textwidth]{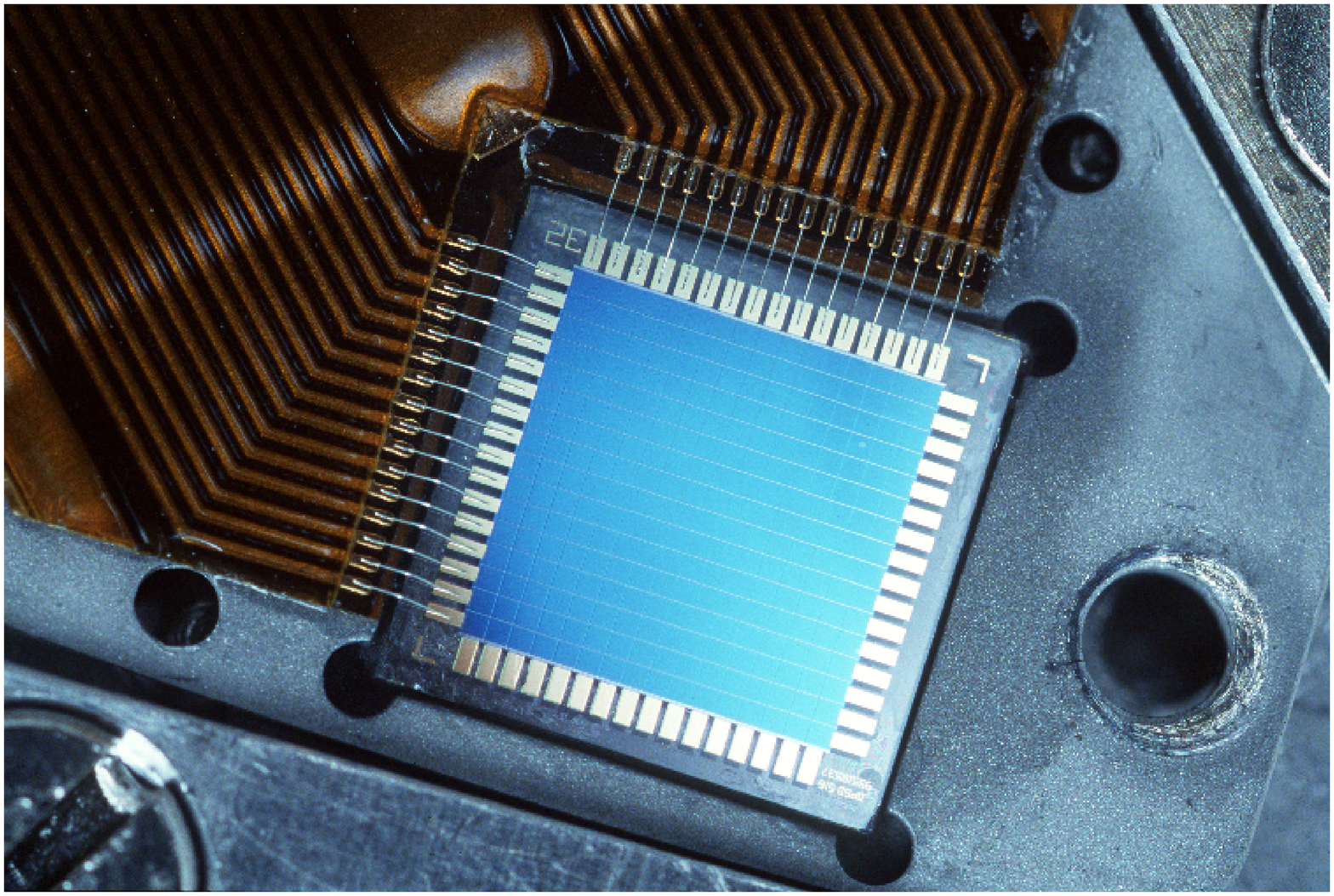} 
       \end{tabular}
        \caption{Top figures: the standard and reversed sensor orientations. Bottom:
    photo of a standard sensor showing the bonding of the flat readout cables.}
        \label{fig:local-xy}
    \end{center}
\end{figure}

\begin{table}[h]
\caption{The positions of the sensor planes along the $z$-coordinate axis 
(parallel to the beam line) in the ZEUS reference frame with the origin
at the nominal electron--proton interaction point. The plane positions
are also shown in Fig.~\ref{fig:la-scheme}.}
\label{tab:sensors-z}
\vskip0.2cm
\begin{center}
\begin{tabular}{|c|c|l|c|c|}
\hline
Index& Plane & & $z$ (mm)   & Comment  \\
\hline
0 & RCTD & Rear CTD end plate& $-1078.1$ & reversed \\
1 & RMVD & Rear MVD end flange& $-1007.6$ & reversed \\
2 & RBarrel & Rear MVD barrel flange& $-360.4$ & standard \\
3 & FBarrel & Forward MVD barrel flange& $ 302.6$ & standard \\
4 & Wheel-1 &  & $ 437.1$ & missing \\
5 & Wheel-3 & MVD wheel-3& $ 717.1$ & standard \\
6 & FMVD & Forward MVD end flange& $ 1104.6$ & standard \\
7 & FCTD & Forward CTD end plate& $ 1179.1$ & standard \\
\hline
\end{tabular}
\end{center}
\end{table}
Sensors are positioned in planes perpendicular to the beamline at seven 
locations:\footnote{Table~\ref{tab:sensors-z} gives the precise positions.}
0 (RCTD) and 7 (FCTD) are attached to the CTD at its rear and forward
end-plates, respectively; 1 (RMVD) and 6 (FMVD) are just inside the rear and forward 
end flanges of the MVD support tube,respectively; 2 (RBarrel) and 3 (FBarrel) are on 
the rear and forward MVD barrel-flanges and 5 (Wheel-3) is at
the position of MVD most forward wheel, wheel-3. 
Plane-4 at the position of wheel-1 could not 
be installed because of space constraints. The TSD sensor
strip signals are read out in a local $x-y$ coordinate system where $x$ is
defined to be from the bottom ohmic contacts (cathodes) and $y$ from the upper bias 
voltage contacts (anodes). For space reasons the sensors on the two planes
at the rear end had to be mounted in a reversed orientation.
Fig.~\ref{fig:local-xy} shows the orientation of
the local coordinate systems for standard and reversed sensors (top figures).
The bottom figure shows the bonding of the special flat readout cable to sensor strips
in the standard mounting. Note that the coordinate
will vary in a direction at right-angles to the readout strips.

The relationship between the sensor local coordinate system 
and the ZEUS coordinate system depends on the sensor location. More details
are given in the section on reconstruction and data analysis. The readout
cables for the sensors presented quite a challenge. Although the MVD has
a much higher spatial precision than the large drift chamber outside it, 
the benefit of these points could be significantly reduced in the overall 
track fit if the MVD increased the multiple scattering by too much. 
Thus the readout cables had to have minimum mass and, as there was no space 
for on- or near-detector amplification, the impedance of the cables over
the total length from sensor to readout board of around $25\,$m had to
be kept low. A special cable was designed and fabricated by Oxford engineers and
technical staff. Using a photo-fabrication technique 
$32\times 250\,\mu$m tracks were etched onto a $1\,$m$\,\times\,0.5\,$m 
flexible kapton-backed copper sheet. Then using an ingenious `cut and fold' 
technique continuous flexible cables with thickness of only $200\,\mu$m and of
various lengths up to $20\,$m were produced.

\subsection{Lasers and optical fibres}

The wavelength of the laser should be long enough to give good sensor
transmission, but short enough to have adequate sensor sensitivity.
The available aperture for the laser beam is  $5\times 5\,$mm$^2$ and the
beam should be contained in the aperture over a length of $2\,$m.  Taking all 
these requirements into consideration, the laser was chosen to have a wavelength of 
$780\,$nm, a gaussian profile and a beam diameter of $1.5\,$mm at the waist.
More details on the characterisation of the laser beam are given in
reference \cite{mvdla1}.  
\begin{figure}
    \begin{center}
        \includegraphics[width=\textwidth]{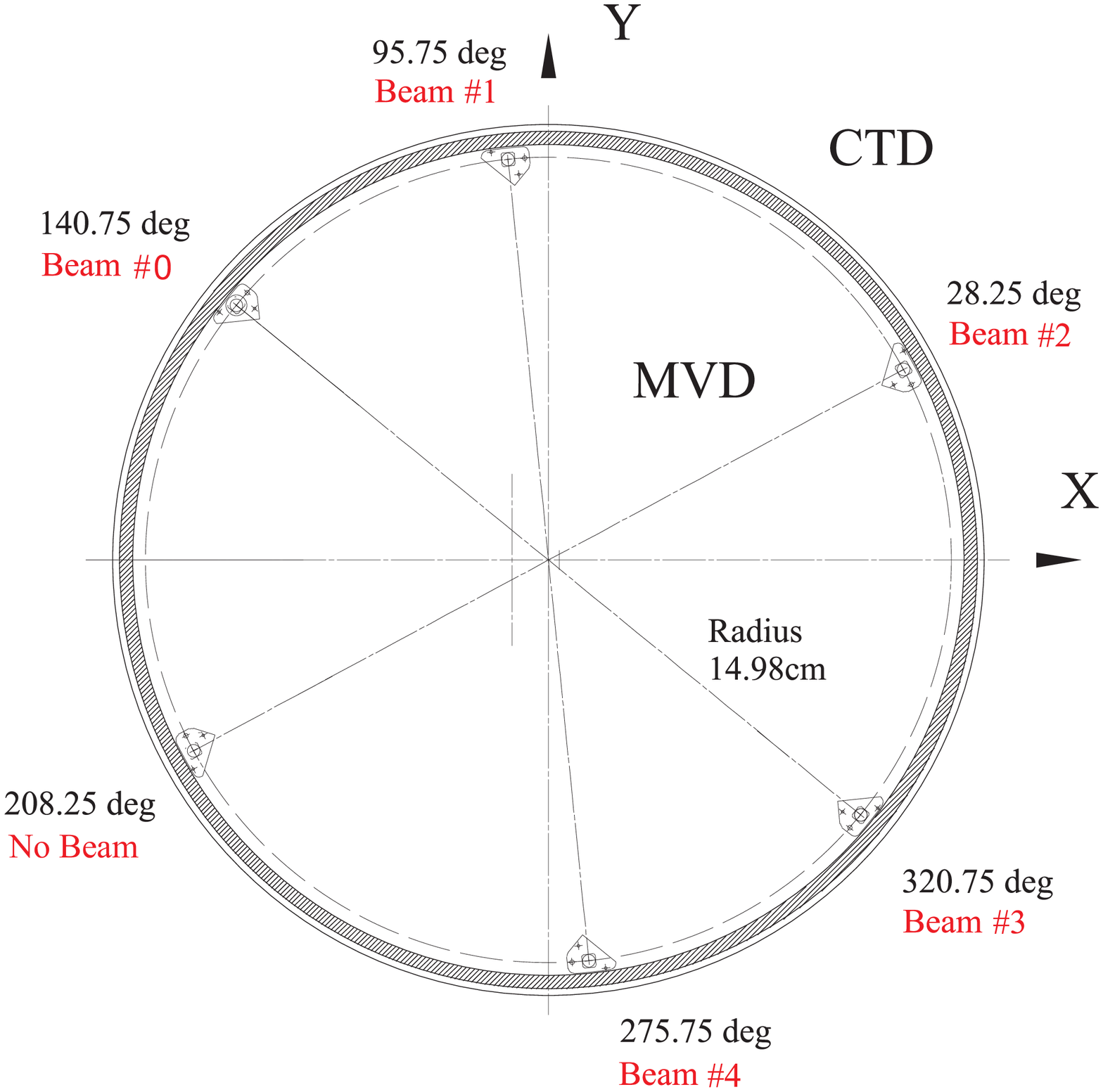}
        \caption{The positions and numbering of the five laser beams around the
circumference of the MVD support tube. View is from the forward end of the CTD in
the ZEUS coordinate system.}
        \label{fig:laser-pos}
    \end{center}
\end{figure}
The positions of the five laser beams around the circumference of the MVD support
tube are shown in Fig.~\ref{fig:laser-pos}. A sixth laser beam was planned at the
position $208.25^\circ$ but had to be abandoned because of conflicting demands
for space. The numbering and power of the
lasers are given in Table~\ref{tab:lasers}. Unfortunately beam-$0$
was knocked out of alignment during work to modify a radiation absorber inside
the beampipe at the rear end of the MVD. Given the extreme difficulty of reaching
the optical fibre laser heads at this time, the beam could not be re-aligned.
\begin{table}[h]
\caption{Laser beam positions, numbering and power. The details are from
measurements made in April 2006. At this time some lasers were moved between beam 
lines, to get the best match between lasers and working sensors.}
\label{tab:lasers}
\vskip0.2cm
\begin{center}
\begin{tabular}{|c|c|c|c|}
\hline
Beam& Position & Power (mW)  & Comment  \\
\hline
0 & $140.75^\circ$ & $2.9$ & lost inside MVD barrel \\
1 & $95.75^\circ$ & $4.7$ &  noisy \\
2 & $28.25^\circ$ & $4.4$ &  good\\
3 & $320.75^\circ$ & $3.0$ &  power low\\
4 & $275.75^\circ$ & $3.7$ & OK \\
5 & $208.25^\circ$ & - & not installed \\
\hline
\end{tabular}
\end{center}
\end{table}
The lasers, IFLEX600 CW with maximum power rating of $5.2\,$mW and $0.65\,$mm
collimated output, were provided by Point Source~\cite{pointsource}.
The lasers and their power supplies are positioned with other MVD service
and readout crates about $25\,$m from the detector. The laser beams are carried 
by optical fibres,\footnote{Point Source $5\,$mm outer diameter stainless-steel 
jacket fibres.}
following the cable route through the patch box, with fibre connectors, to the laser 
fibre heads at the rear end of the MVD support tube. The fibre heads are mounted 
on a ring attached to the CTD and `pointing' adjustments for a laser beam are made 
using three screws for each head. Given the very limited space between the beam pipe 
and the CTD rear on-board electronics, the final adjustment is a difficult and 
delicate task, that becomes impossible once the rear tracking detector is installed. 

The lasers and their power supplies are contained in two `shutter boxes', one box with 
three lasers and the other box with two lasers.
Once the lasers are powered on, the light output can be `switched' on or off by 
opening or closing a mechanical shutter across all three laser beams. The shutter is
moved by an electromagnet which can be under operator or software control. The shutter
control is also interfaced to the ZEUS slow-control and safety interlock systems.
In normal operation the lasers are powered continuously to avoid fluctuations when they
are powered up. The beams are switched on or off for the various data collection 
procedures by use of the shutters.

\subsection{Sensor Resolution}
Before installation the TSD sensors were characterised using a test set-up 
consisting of a computer controlled two-dimensional translation stage mounted 
on an optical rail system~\cite{mvdla1}. For a given sensor plane the basic 
measurements are strip signals above the pedestal values. For a laser beam
reasonably well-centred on the sensor and perpendicular to the sensor plane,
the strip signals follow a gaussian distribution. Full details on how
the position measurements are made in the final system are given in 
Section~\ref{sec:recons}. 

As originally noted by Kroha~\cite{kroha2},
the resolution for measurement of the laser beam position could be improved
by correcting for variations in the thickness of the amorphous silicon layer and
the glass substrate. Such variations may lead to interference patterns in
the transmission and absorption of the laser light thus affecting the beam
position measurement, see also Bauer et al~\cite{bauer}.
Correction matrices with a $250\,\mu$m grid were measured for all sensors. 
Using the correction matrices, the best resolution obtained in the test setup was
less than $2\,\mu$m -- details are given in reference \cite{mvdla1}. However, 
it was decided not to use the correction matrices in the routine analysis 
of the laser data as other systematic effects were much larger. These will
be discussed in Section~\ref{sec:recons}.  

%\input readout
%%
%% readout
%%
\section{Readout and control}\label{sec:readout}

The photo-current signals from the sensor strips are carried from the MVD by
the special cables described in Section~\ref{ssec:sens-cab} via the patch-boxes 
to the readout cards. For each of the eight $z$-locations along the MVD support 
tube all five TSD sensors are read out by a single card and all the cards are
located in a VME crate at the MVD services area. There the signals are 
multiplexed, amplified by a 
current-voltage  transformer, digitised and stored in memory on VMEbus boards. 
This memory is addressed using a complete rewrite of the VME driver to exploit 
Motorola's TUNDRA chipset \cite{uvmelib,vmedaq}. The data are transferred via 
TCP/IP and made available to the ZEUS event building components.

The laser alignment slow control and data acquisition are fully integrated into the 
existing ZEUS MVD framework \cite{zeusmvd2}, which is based on VME board 
computers running Lynx OS and Intel PCs running SuSE GNU/Linux. This slow 
control incorporates an interface to the ZEUS safety control system prohibiting
operation of the lasers when the beam lines are accessible to people.

The readout and slow control systems have been designed to allow two modes of
operation \cite{mvdla-control}. For the first the laser system is fully integrated 
with the 
ZEUS run control system, data taken is stored on tape as part of the main ZEUS 
data store. It is then available for analysis along with the slow-control and
general environmental records from normal data-taking. For the second mode the 
system can run in parallel with normal ZEUS data taking, but the laser data is 
now stored on disks on the Intel PC's via NFS. ASCII format 
copies of laser data from both data-taking modes are also stored on a local 
MVD computer disk.  

%\input recons
%% data-basics
%%
\section{Data analysis and first results}\label{sec:recons}

The raw data from a single laser run comprises two sets of ADC counts for both local
$x$- and $y$-coordinate strips at each sensor plane along all five beam lines. The
first set is taken with the laser shutters closed to establish the pedestal values, 
followed by the second set with shutters open. The raw signals are then given by
subtracting the pedestal values from the `laser-on' values.
Early studies of the laser alignment data showed that typical values for the dark
currents of the anode and cathode strips were around 0 and up to 50 ADC counts,
respectively. A number of cuts and corrections were applied to strip signals, $I_i$, 
after pedestal subtraction and before applying the position algorithm:
\begin{itemize}
\item
negative strip currents ($I_i<0$) set to zero;
\item
isolated `hot strips' with $I_i>1000\,$counts set to zero;
\item
strips with large ratios to both neighbours 
($I_i/I_{i+1}~{\rm and}~I_i/I_{i-1}>10$) rejected;
\item 
`empty' sensor plane with $\sum_{i=1}^{16}I_i<160\,$counts rejected;
\item
to avoid edge effects the first two and last two strips are ignored.
\end{itemize}

\subsection{Mean position}

\begin{figure}
    \begin{center}
        \includegraphics[width=\textwidth]{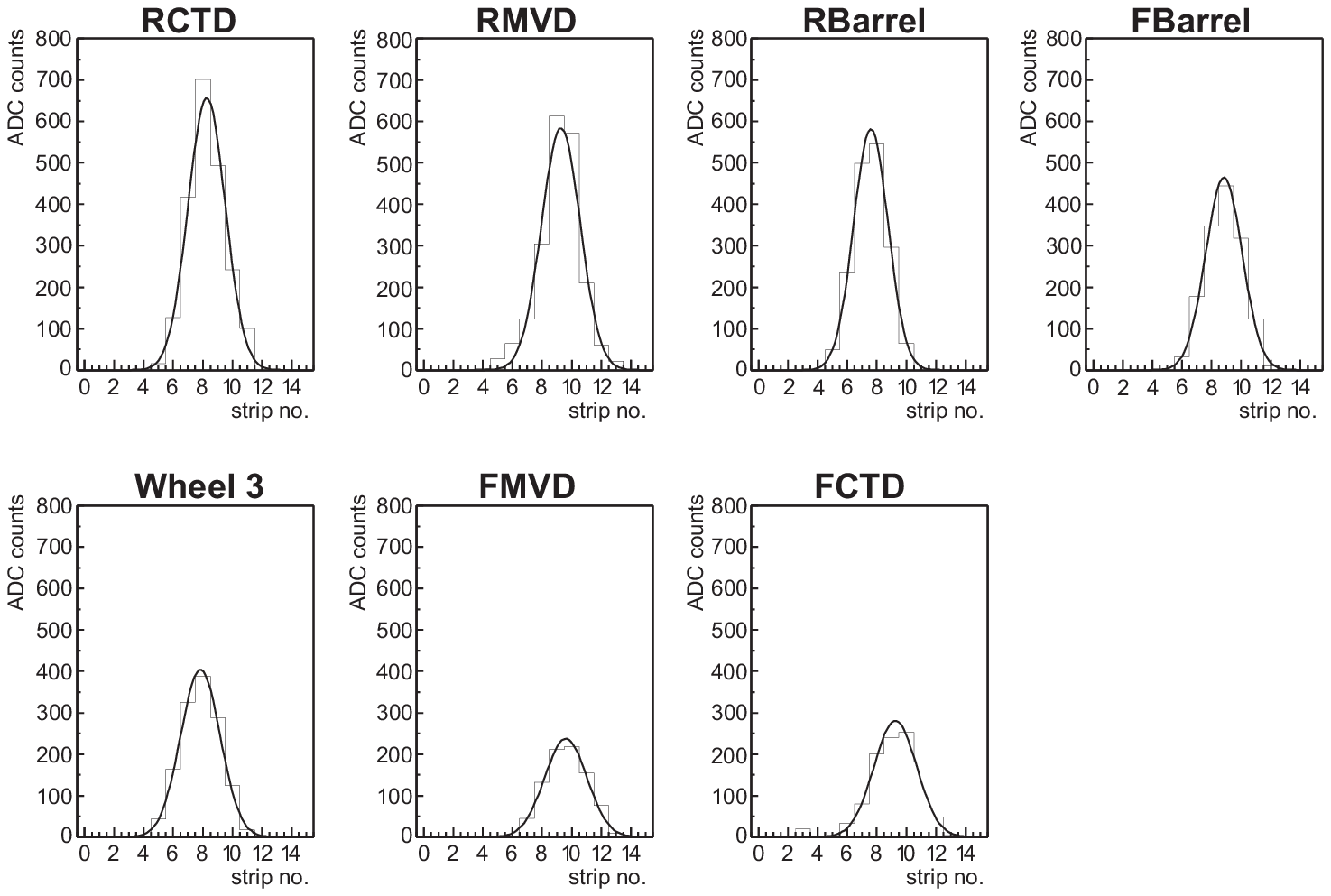}
        \caption{Examples of the beam-profile signals from the sensors along one beam 
line, together with the gaussian fits. The strip currents are in units of ADC counts 
and are plotted as functions of strip number. RCTD corresponds to sensor plane 0 and
FCTD to sensor plane 7 -- see Table~\ref{tab:sensors-z} for details.} 
        \label{fig:beamprofs-vs-z}
    \end{center}
\end{figure}

Two methods for determining the mean position, $\bar{x}$, and resolution, $\sigma$, of 
the signal in a sensor 
plane were considered. The simplest is to use a current weighted mean
$$\bar{x}=\sum_ix_iw_i~~{\rm with}~~w_i=\frac{I_i}{\sum_i I_i}~~{\rm and}~~
\sigma=\frac{d}{\sqrt{12}}\sqrt{\sum_i w_i^2},$$ 
where $d\approx 300\,\mu$m is the strip width. The strip position, $x_i$, is
taken to be at the centre of the strip and the error is assumed to be uniformly 
distributed over the strip width. This gives a conservative value of around
$40\,\mu$m for the mean position resolution. A more sophisticated approach is to
assume a gaussian profile for the strip-current distribution of a sensor plane. The
parameters $\bar{x}$ and $\sigma$ are determined by fitting the profile to
$\displaystyle N\exp\left[-(x-\bar{x})^2/(2\sigma^2)\right]$, with the error
on an individual strip position taken to be $1/\sqrt{I_i}$. The fit is performed
twice, with the second fit scaled to enforce $\chi^2/{\rm ndf}=1$.  
The typical resolution from the fit method is below $10\,\mu$m. In more detail, an 
analysis of about 2580 position measurements showed that 95\% of the errors were below 
$15\,\mu$m, with a mean of $6.5\,\mu$m, with the remaining 5\% of errors extending
up to $40\,\mu$m. The gaussian fit method is the one used for all the results 
described below. Fig.~\ref{fig:beamprofs-vs-z} shows the beam-profile
signals and gaussian fits along one beam line. The attenuation of the laser intensity
and beam broadening are evident as the number of sensors traversed increases.
The attenuation is roughly consistent with the 80\% transmission found in the test
system.

\begin{figure}
    \begin{center}
         \includegraphics[width=0.75\textwidth]{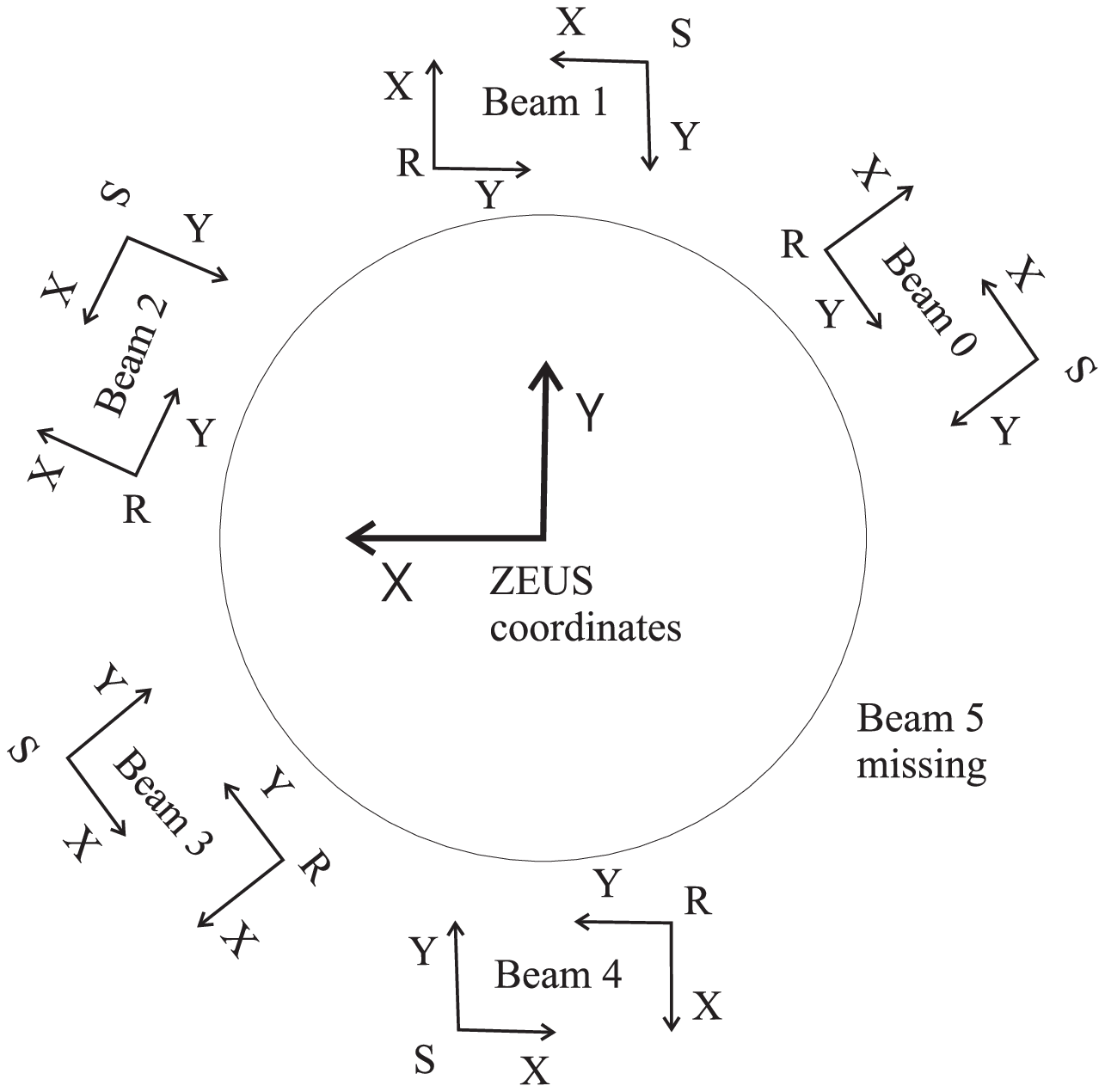}
        \caption{The orientation of sensor local coordinates with respect to
the ZEUS coordinate system, viewed from the rear CTD end. The labels S and R refer 
to standard and reversed sensors, respectively.} 
        \label{fig:localxy}
    \end{center}
\end{figure}

The mean positions may be plotted in sensor local
coordinates or in ZEUS coordinates. The geometrical relationship between them is shown 
Fig.~\ref{fig:localxy} and the transformation from the local system, 
${\bf r}_{local}$,
to the common ZEUS system, ${\bf r}_{ZEUS}$ may be written as
$${\bf r}_{ZEUS} = {\mathcal R}\cdot{\bf r}_{local} + {\mathcal T}$$
where ${\mathcal R}$ is a $2\times 2$ rotation matrix and $\mathcal T$ a 2-D 
translation. 
The transformations also take into account the standard or reversed orientation of 
the sensors.

\begin{figure}
    \begin{center}
         \includegraphics[width=\textwidth]{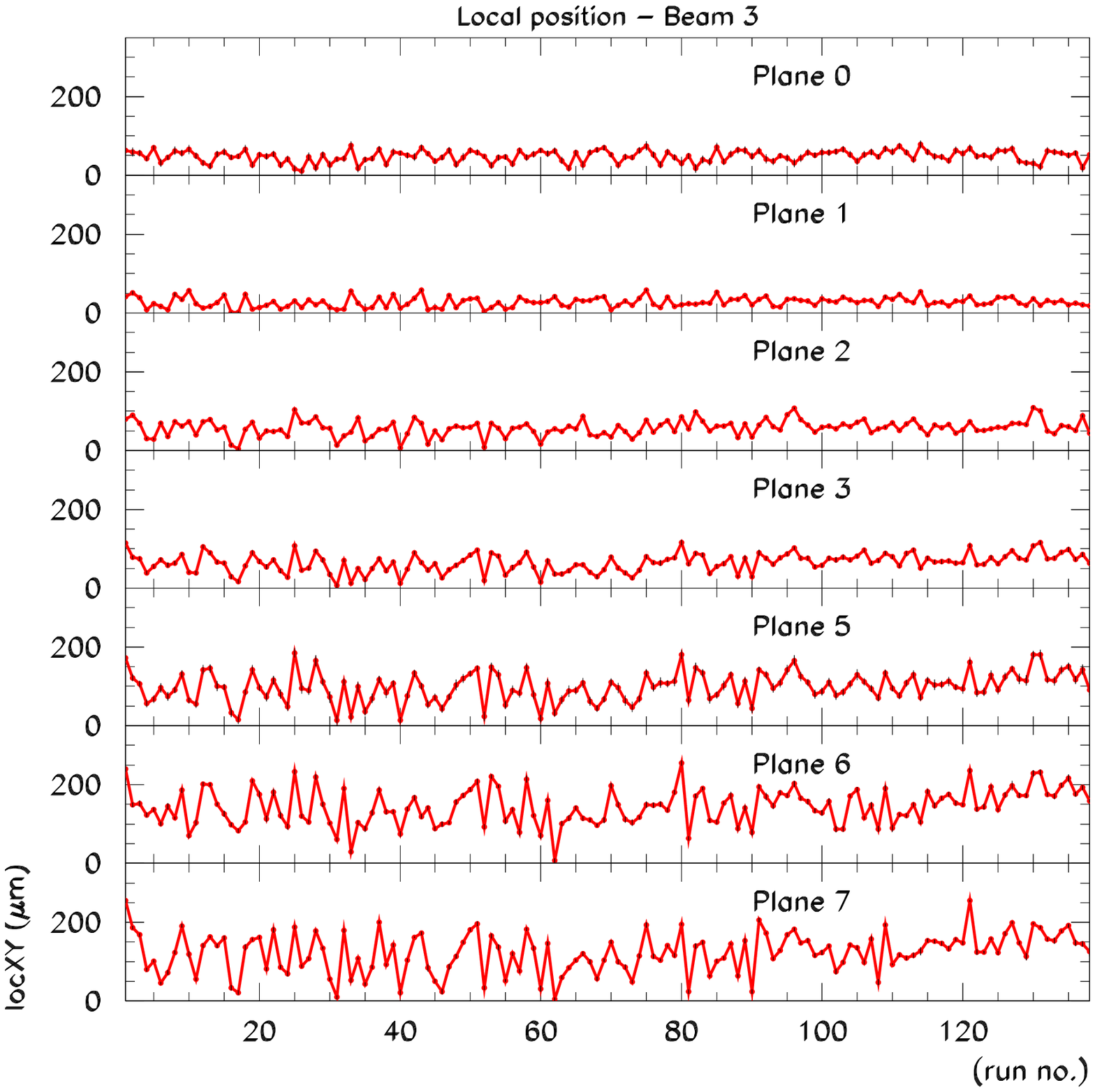}
        \caption{Relative positions of laser beam-3, in local coordinates, at 
the seven planes starting with plane-0 at the top. The $y$ (anode) signals are 
shown as functions of the run number. At the time of these measurements the 
laser at beam-3 was that with the highest power.} 
        \label{fig:rel-beam3-04}
    \end{center}
\end{figure}

During the period January to August 2004, laser data were collected regularly at the 
end of each HERA fill for normal e-p interactions, giving a total of 200 runs. The
first attempt at analysis compared the mean sensor positions of a given laser beam 
relative to a reference run -- usually the first run of the period under study.
The reason for plotting positions relative to a reference run is to allow the data
from different beams to be plotted using a common scale. 
An example of the data for the whole period is shown in Fig.~\ref{fig:rel-beam3-04}. 

The figure shows the relative mean position of the $y$, anode, signal along 
laser beam-3 in each of the sensor 
planes, with plane 0 (RCTD) at the top and plane 7 (FCTD) at the bottom. A number of
points may be made:
\begin{itemize}
\item
the size of deviations tends to increase with distance along the laser beam;
\item
there are some quite large deviations in planes 5, 6, 7, up to $100\,\mu$m or so;
\item
there are clear correlations between the larger deviations in different planes. 
\end{itemize} 
Although not shown, the local $x$-coordinate (cathode) signals shows similar
features but with smaller fluctuations.
The difficult question to answer from this type of plot is whether this is evidence for
movement of the MVD support tube or simply instabilities of the lasers and noise in
the sensors. The data shown in the figure were collected from laser runs separated by
quite long time intervals -- hours or even days between consecutive runs. 
In between laser runs, the lasers were switched off. The lasers were switched on for
a short while before an alignment data run, but there was insufficient time to ensure 
that the system had stabilised because of the exigencies of physics data-taking.

\begin{figure}
    \begin{center}
         \includegraphics[width=\textwidth,angle=270.,scale=0.275]{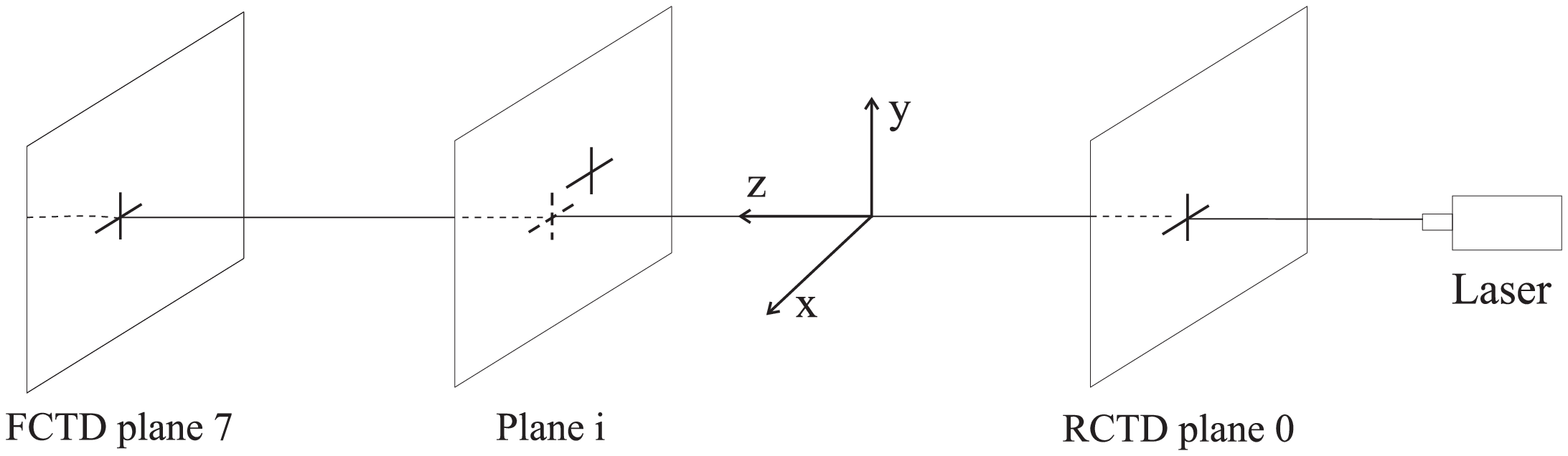}
        \caption{The procedure used to define the reference line for a given laser
beam using the CTD planes 0 and 7 and residual (offset) between the expected position
and measured mean position in an interior MVD sensor plane.} 
        \label{fig:residual}
    \end{center}
\end{figure}

\begin{figure}
    \begin{center}
         \includegraphics[width=\textwidth]{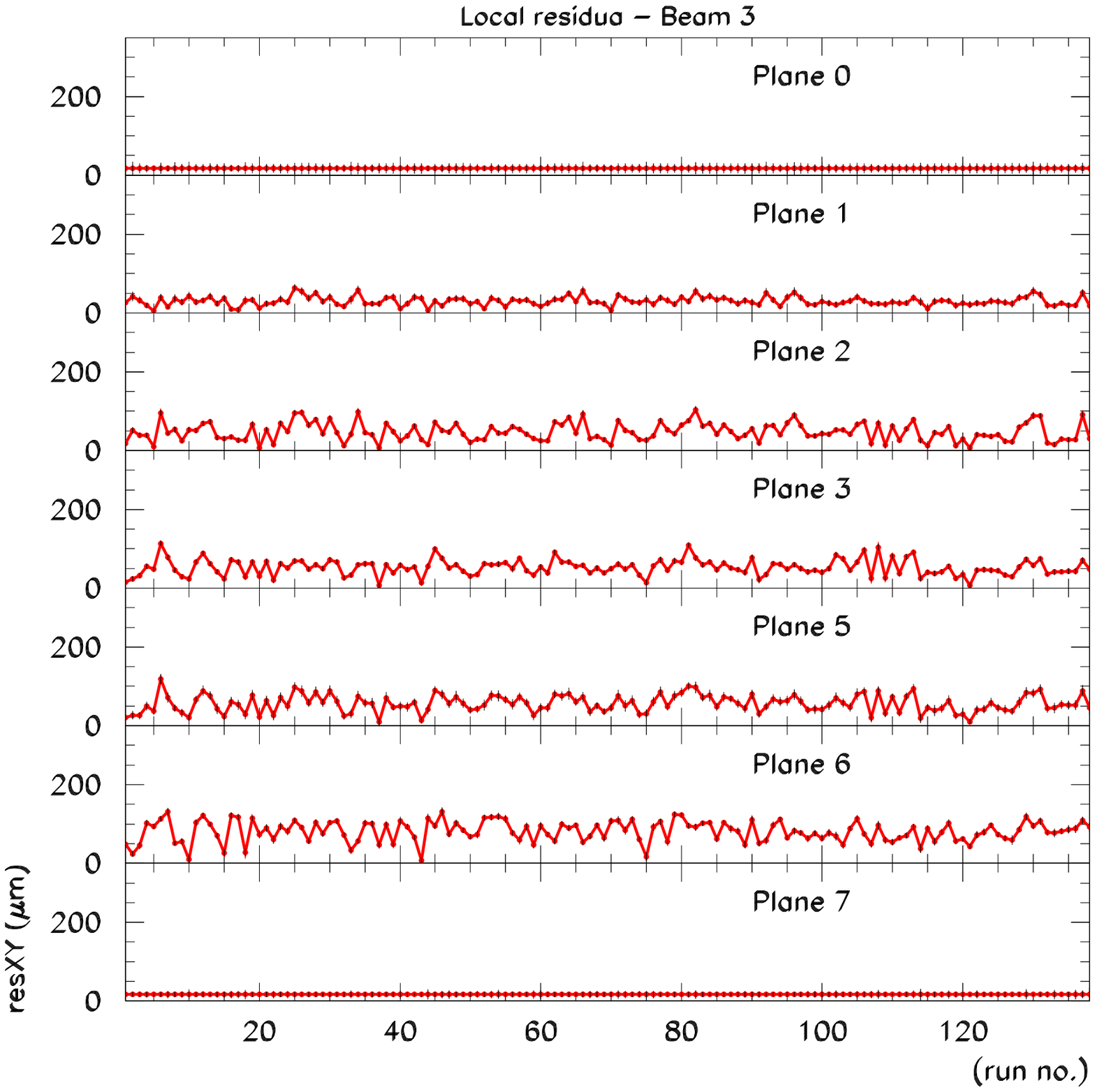}
        \caption{Residuals of laser beam-3, in local coordinates, with
respect to the straightness monitor defined by the laser-beam positions in planes
0 and 7 (R and FCTD). Other details the same as in Fig.~\ref{fig:rel-beam3-04}.} 
        \label{fig:resid-beam3-04}
    \end{center}
\end{figure}

For these reasons an alternative procedure was developed for the analysis of the
laser beam position data. The idea, to define each beam line as an independent
`straightness monitor', is shown schematically in Fig.~\ref{fig:residual}.
For a given laser beam and local coordinate the mean positions in sensor 
planes 0 and 7, attached to the CTD, are used to define the reference straight line.
The expected position of the beam at a sensor plane, $i$, within the MVD, is then
given in local sensor coordinates by:
$$
x_i=x_0\frac{z_i-z_7}{z_0-z_7}+x_7\frac{z_0-z_i}{z_0-z_7},~~~~~
y_i=y_0\frac{z_i-z_7}{z_0-z_7}+y_7\frac{z_0-z_i}{z_0-z_7}. 
$$
The {\it residuals} in the two coordinate directions are calculated as the differences
between the expected positions and the corresponding measured mean
positions, $(x_i-\bar{x}_i^{meas.}),~(y_i- \bar{y}_i^{meas.})$.    
Fig.~\ref{fig:resid-beam3-04} shows the same data as displayed in 
Fig.~\ref{fig:rel-beam3-04}, but now the residuals are plotted relative to the
line defined by planes 0 and 7 (hence the absence of deviations at these positions).
The residuals are plotted relative to those from a reference run, which is chosen to
be the same as that used for these data before. Comparing 
Figs~\ref{fig:resid-beam3-04}
and \ref{fig:rel-beam3-04}, one sees that the fluctuations are smaller, 
particularly in the planes furthest from the optical fibre laser heads.
There is evidence for movement or for further instabilities of the lasers. More
information is clearly needed. 

There is quite a variation in the `noise' level of individual laser beams, 
fluctuations are caused by pointing instability of the source and beam 
deflections from thermal gradients. Beam-3 shown in Figures~\ref{fig:rel-beam3-04} 
and \ref{fig:resid-beam3-04} 
is relatively quiet. Allowing for this variation, similar features as
described above are seen for the other beams in the system. Although there may be 
evidence for some movement, there is no evidence for any permanent `step changes'
in the position of the MVD support tube over the nearly nine months of data considered
in this section. If the residuals for a given beam and sensor are plotted for the 
whole period, the resulting distributions are reasonably gaussian with values of
$\sigma$ around $10-20\,\mu$m for a `quiet' laser beam and around $40-50\,\mu$m for
a `noisy' beam.

%\input results
%%
%% results
%%
\section{Correlation studies}\label{sec:results}

As discussed in the preceding section, to establish whether the variations in
mean position (residual) of a sensor are caused by motion, more information is
needed. The first attempts to provide precision alignment constants for the MVD
tracking sensors used events with through-going cosmic ray tracks only. There
were two sources for the data sets: the first from dedicated cosmic-ray runs
with a special trigger taken when the HERA collider was not operational and the
second from cosmic-ray events selected from the normal e-p interaction data
stream with HERA operational. It was expected that the two types of data would
give the same results (provided that the data were collected during roughly the same
period of time and that no changes were made to the detector configuration). 
They did not and no `trivial' explanations could be found for the differences
in alignment constants which were of magnitude $50-100\,\mu$m.
It was eventually realised\footnote{We thank Drs R. Carlin and U. Koetz for making
this suggestion.}
 that there was a difference in the environmental conditions:
whether the HERA collider magnets were powered on or not. 
As described in Section~\ref{sec:mvd}, the final focusing GG magnet in the HERA-II
configuration reaches well within the detector and quite close to the central 
tracking system. In addition, the MVD readout and services cables are tightly 
wrapped around it. The way the magnet is supported within the detector is
by a strap from above and this would allow some movement, indeed position sensors
attached to the magnet showed that it does move slightly.

Two further steps were required before this idea could be tested quantitatively: 
the first was to change the mode of operation of the laser system and the second was to
get access to the temporal records of the GG magnet current. To reduce laser
instabilities from switching the lasers on and off, it was decided to leave the
lasers on permanently (while the detector was closed for data taking) and control
laser runs by the mechanical shutter. 
With this change laser runs could be taken at much shorter
intervals. To study possible external effects, the laser data
were collected every 4 minutes for periods of hours, up to a maximum
of nine days. These data sets gave ample opportunity for study of the effects of both 
regular and irregular operation of the ZEUS detector.  
The circulation of bunches in HERA provides a very accurate
clock signal that is used for synchronisation both within the ZEUS
experiment and between the experiment and HERA, so it was relatively straightforward
to relate the state of the GG magnet current to the timing of laser runs.  

Other sources of environmental change were also considered, such as the temperature
of the MVD, the temperature of the beam pipe and the temperature of the CTD. The
last two were quickly ruled out, but it was found that the temperature within the MVD
could change by almost 10$^\circ$C depending on whether the onboard electronics were
powered or not.

\begin{figure}
    \begin{center}
         \includegraphics[width=\textwidth]{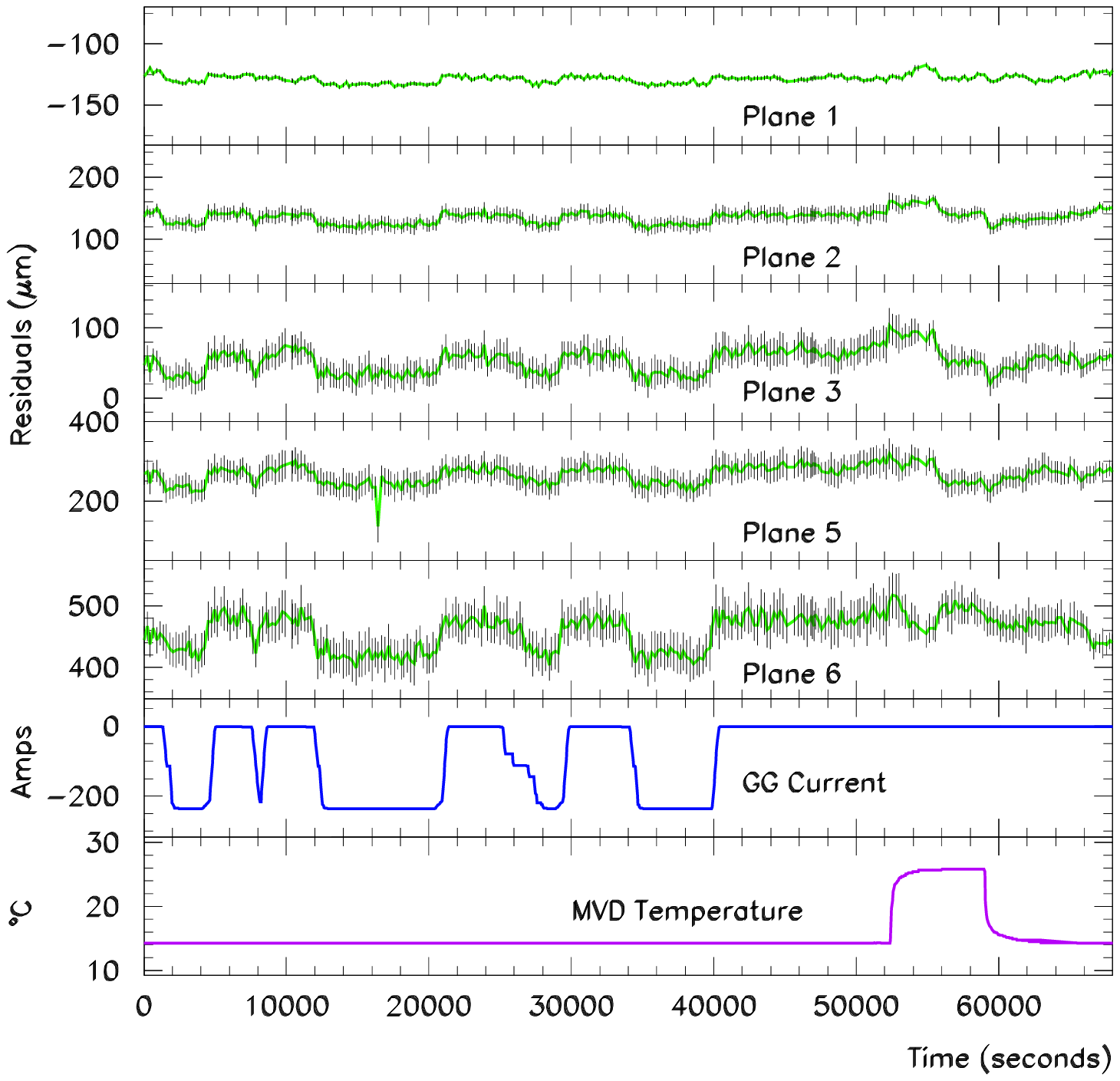}
        \caption{Local-$x$ (cathode) residuals for beam-2, planes 1, 2, 3, 5 \& 6 
with the GG magnet current and the MVD temperature in the bottom two plots, all 
as a function of elapsed time.} 
        \label{fig:cor3000-local-x}
    \end{center}
\end{figure}

Fig.~\ref{fig:cor3000-local-x} shows an example of the results from the correlation
studies. It shows the local-$x$ (cathode) coordinate residuals for laser beam-2 at the 
five planes within the MVD 
(upper five plots) together with the GG magnet current and the MVD temperature in
the lowest two plots. All are plotted against the common elapsed time synchronised
by the HERA clock. Note that the residuals are the actual values, relative to the
beam-2 laser line, at each measuring plane and not relative to a reference run.
The temperature in the MVD is measured at a position near wheel-3.
\begin{figure}
    \begin{center}
         \includegraphics[width=\textwidth]{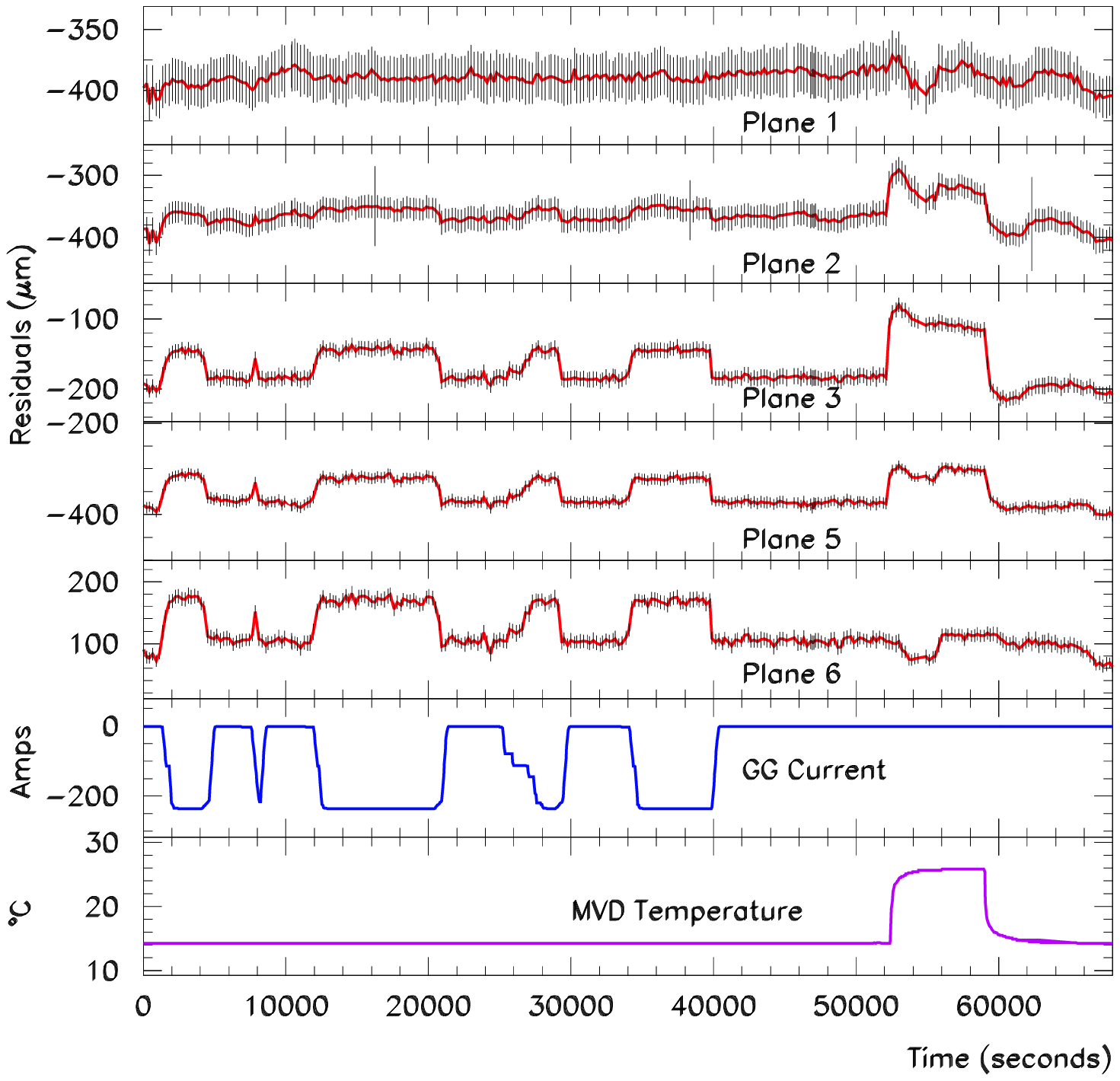}
        \caption{Local-$y$ (anode) residuals for beam-2, planes 1, 2, 3, 5 \& 6 with 
the GG magnet
current and the MVD temperature in the bottom two plots, all as a function of
elapsed time.} 
        \label{fig:cor3000-local-y}
    \end{center}
\end{figure}
Fig.~\ref{fig:cor3000-local-y} shows a similar plot from the same runs for the
local-$y$ (anode) coordinate residuals.

Considering the magnet current first, Fig.~\ref{fig:cor3000-local-x} (local-$x$)
shows that between times of 0 and $45000\,$s there is a correlation in time 
between the mean residual values themselves and with the magnet current being zero or
non-zero. There is also a tendency for the size of the movement to increase with 
increasing plane index. Fig.~\ref{fig:cor3000-local-y} (local-$y$)
shows a similar, but more pronounced correlated movement. This pattern of
movement is also seen in the data from other laser beams and is consistent with the 
MVD support tube being tilted about a fulcrum near the rear CTD attachment. The
assumption is that the MVD support tube moves when the GG magnet moves, via the
MVD cables. 

Regarding correlated movement with temperature, the evidence is very clearly seen in
the local-$y$ plots (Fig.~\ref{fig:cor3000-local-y}) for times between 50000 and 
$65000\,$s. The changes in temperature occurs when the MVD electronics are switched on 
or off. The pattern of movement is different in detail to that related to the GG 
magnet motion, the biggest effects are now seen in 
the three sensor planes nearest to the on-detector electronics, at planes 2, 3 and 
5 (MVD barrel flanges and wheel-3). Fig.~\ref{fig:cor3000-local-x} -- local-$x$
coordinate for beam-2 -- shows a much less clear correlation. The concentration of
movement at these three sensor planes is also seen in other beam line data.
\begin{figure}
    \begin{center}
         \includegraphics[width=\textwidth]{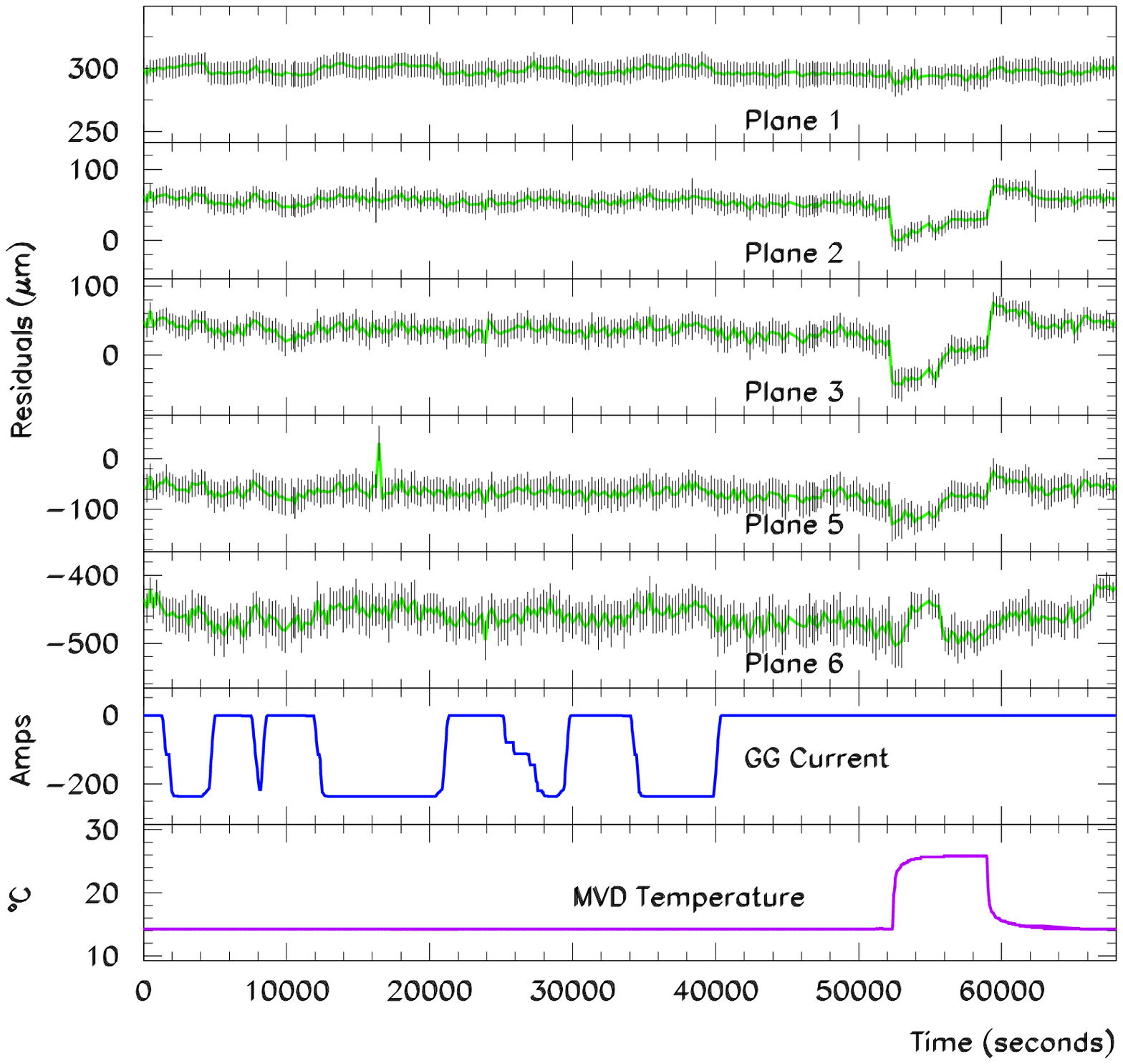}
        \caption{ZEUS frame $x$-residuals for beam-2, planes 1, 2, 3, 5 \& 6 with the 
GG magnet current and the MVD temperature in the bottom two plots, all as a function of
elapsed time.} 
        \label{fig:cor3000-global-x}
    \end{center}
\end{figure}

\begin{figure}
    \begin{center}
         \includegraphics[width=\textwidth]{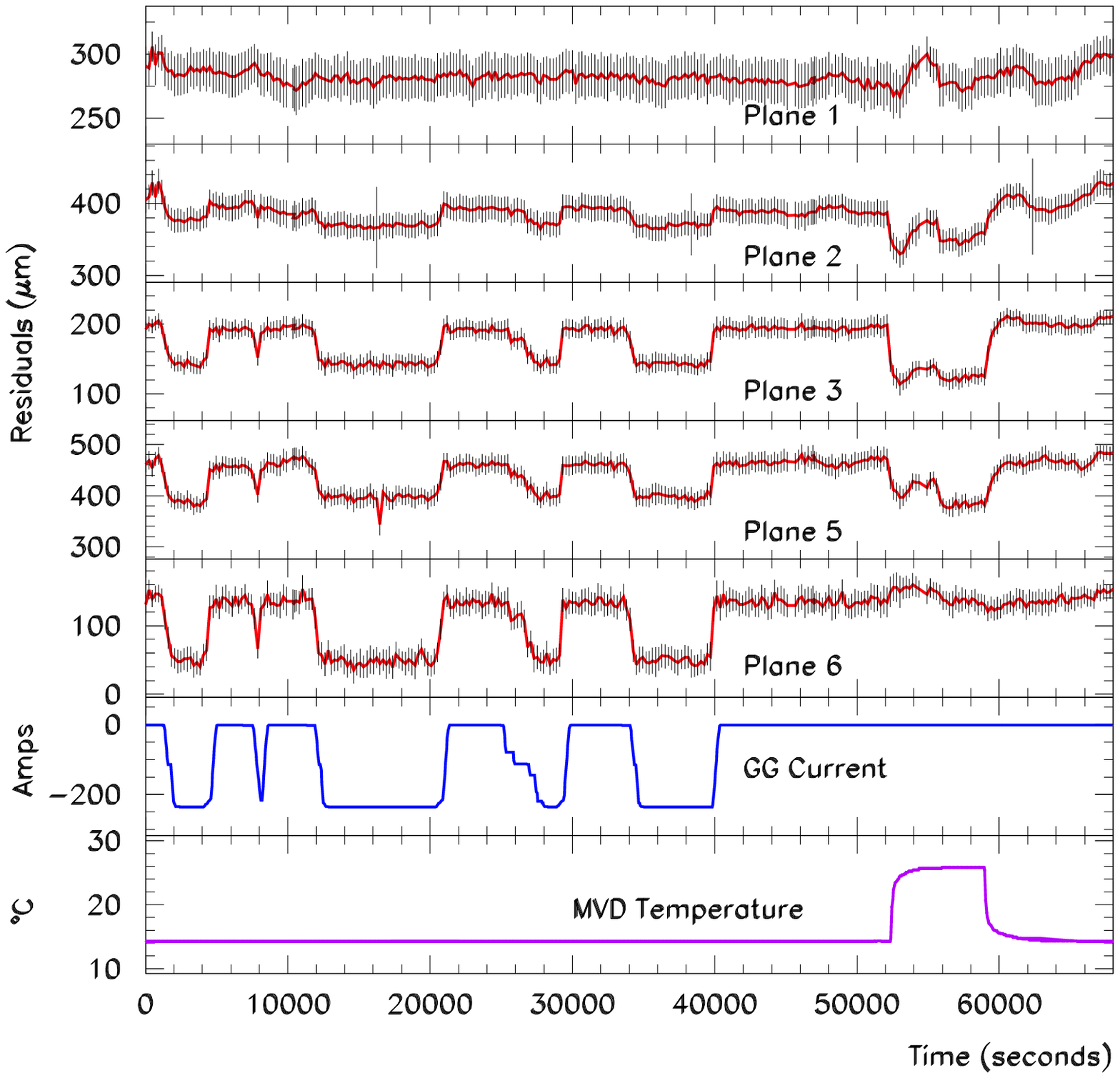}
        \caption{ZEUS frame $y$-residuals for beam-2, planes 1, 2, 3, 5 \& 6 with the 
GG magnet current and the MVD temperature in the bottom two plots, all as a function of
elapsed time.} 
        \label{fig:cor3000-global-y}
    \end{center}
\end{figure}

Figs~\ref{fig:cor3000-global-x} and \ref{fig:cor3000-global-y} show the same beam-2
data but now the residuals have been transformed to the global ZEUS $x$- and 
$y$-coordinates, respectively. At this beam position the GG magnet associated 
movement is 
mainly along the ZEUS $y$-axis, whereas the MVD temperature movement is seen in both 
global $x$ and $y$ directions. Originally it was thought that one might be able
to deduce the nature of the movement of the support tube by fitting the pattern of
local movements to changes expected from a set of `standard motions' -- for example
twists and sags. This might well have been unrealistic even with the full production
system as designed, but it is impossible with reduced number of beam lines and reliable
sensors. However the system is able to give some quantitative information on the
magnitude of local movements and to track such changes.

\begin{figure}
    \begin{center}
     \includegraphics[width=\textwidth]{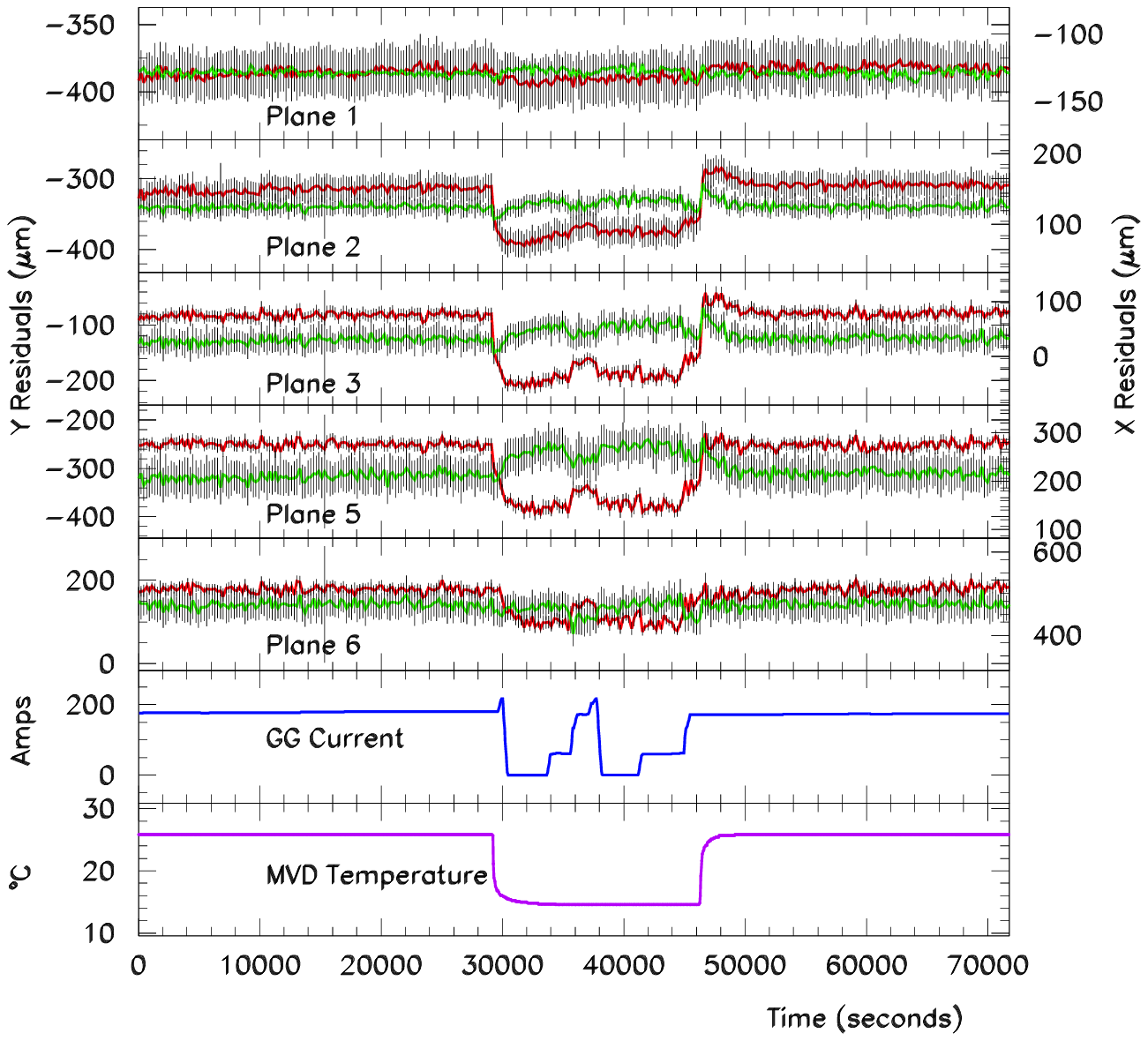}
        \caption{Residuals for local-$y$ (dark grey left-hand scales) and local-$x$
(light grey right-hand scales) are shown for two periods of e-p data-taking with 
MVD and GG magnet both on and a period in between when the MVD was off with the 
GG current being varied with two short sub-periods when it was also off.} 
        \label{fig:stable-ep-run}
    \end{center}
\end{figure}

The residual plots indicate that the position of a sensor is stable 
while the external conditions are stable and that the positions of stability are
themselves stable and reproducible. This has been investigated in more detail by 
averaging the residuals for periods of stability during long laser runs. The periods 
are defined by external changes, but regions of rapid change are excluded. Results
from a typical long period of runs are shown in Fig.\ref{fig:stable-ep-run}. Two long
periods, corresponding to e-p data-taking when both the MVD and GG magnet are on
can be seen. In between is a period of beam injection and acceleration when the MVD is
off throughout and the GG magnet current is varying, but has two short periods 
when it is also off. This pattern is typical and shows that `off' periods are shorter
than those when everything is on. 

\begin{figure}
    \begin{center}
         \includegraphics[width=\textwidth]{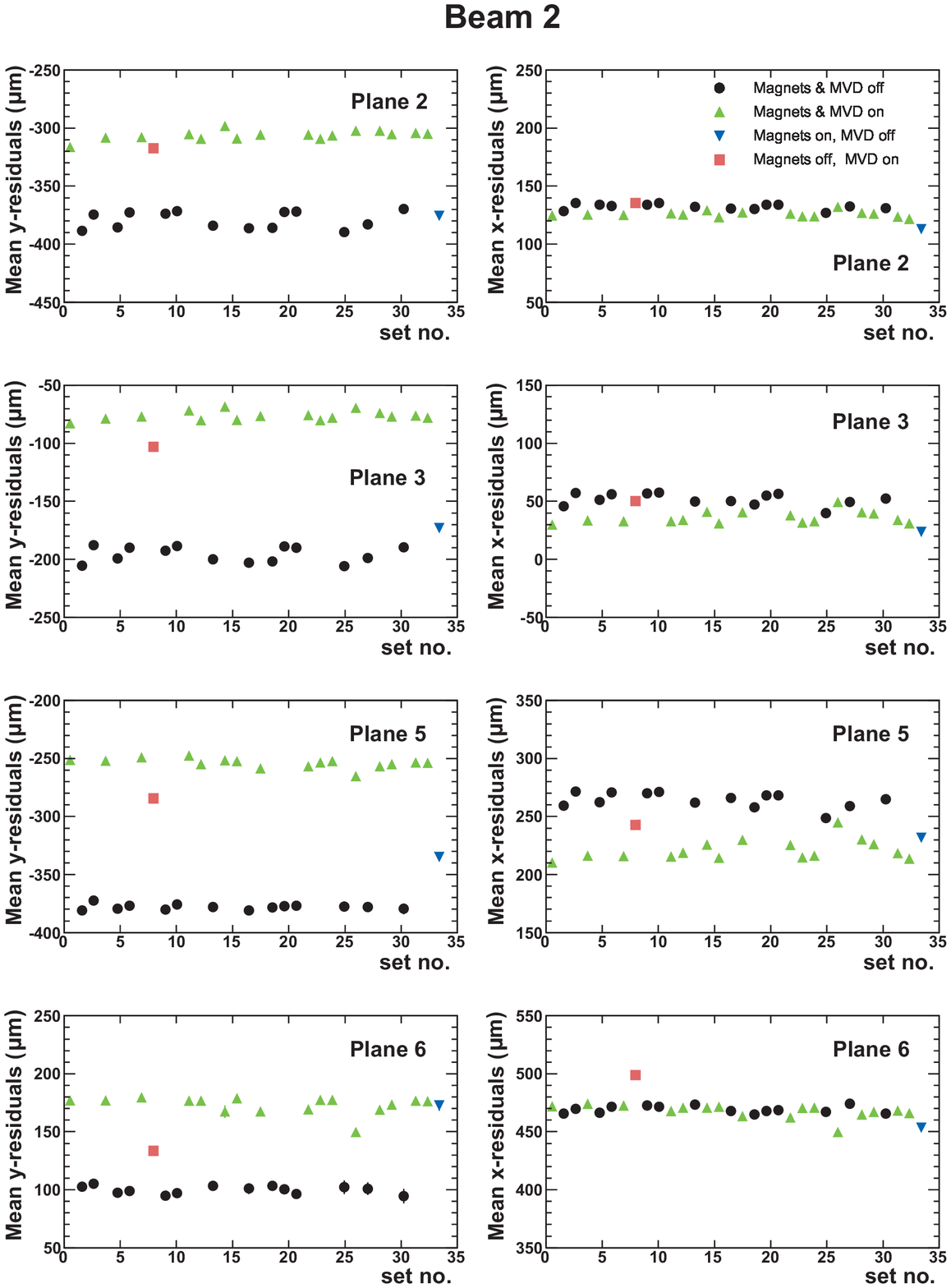}
        \caption{Mean values of residuals for local-$y$ (left-hand plots) and local-$x$ 
(right-hand plots) for beam-2 during periods of
stability. The symbols indicate different conditions: triangle MVD and GG magnet 
both on; circle MVD and GG both off; square MVD on, GG off ($8^{\rm th}$ data set only); 
inverted triangle MVD off, GG on (last data set in the sequence). The size
of the standard deviation is shown by the vertical error bar.} 
        \label{fig:mean-res-local}
    \end{center}
\end{figure}

The detailed results depend on the noise quality of lasers and the sensors. 
For beams 2, 3 and 4 most residual means have RMS values better than $10\,\mu$m, 
beam 1 is noisier and the local-$x$ residuals at  plane-6 appear to be unstable.
The positions for normal running with both the MVD and GG magnet powered and
those with both MVD and GG magnets off are of particular interest. 
The results for local coordinates and beam 2 are shown in 
Fig.~\ref{fig:mean-res-local}. 
Each point shows the mean and standard deviation (shown as a vertical error bar, 
often smaller than the symbol size) for a data set corresponding to 
a single period of stability. The means
for the MVD and GG magnet both on are shown as triangles and when both are off 
as circles.
The data show clearly that the structure is moving between two well-defined positions 
`ON' and `OFF'. The movement is mainly in the local-$y$ coordinate and can be as
much as $120\,\mu$m.   

\begin{figure}
    \begin{center}
         \includegraphics[width=\textwidth]{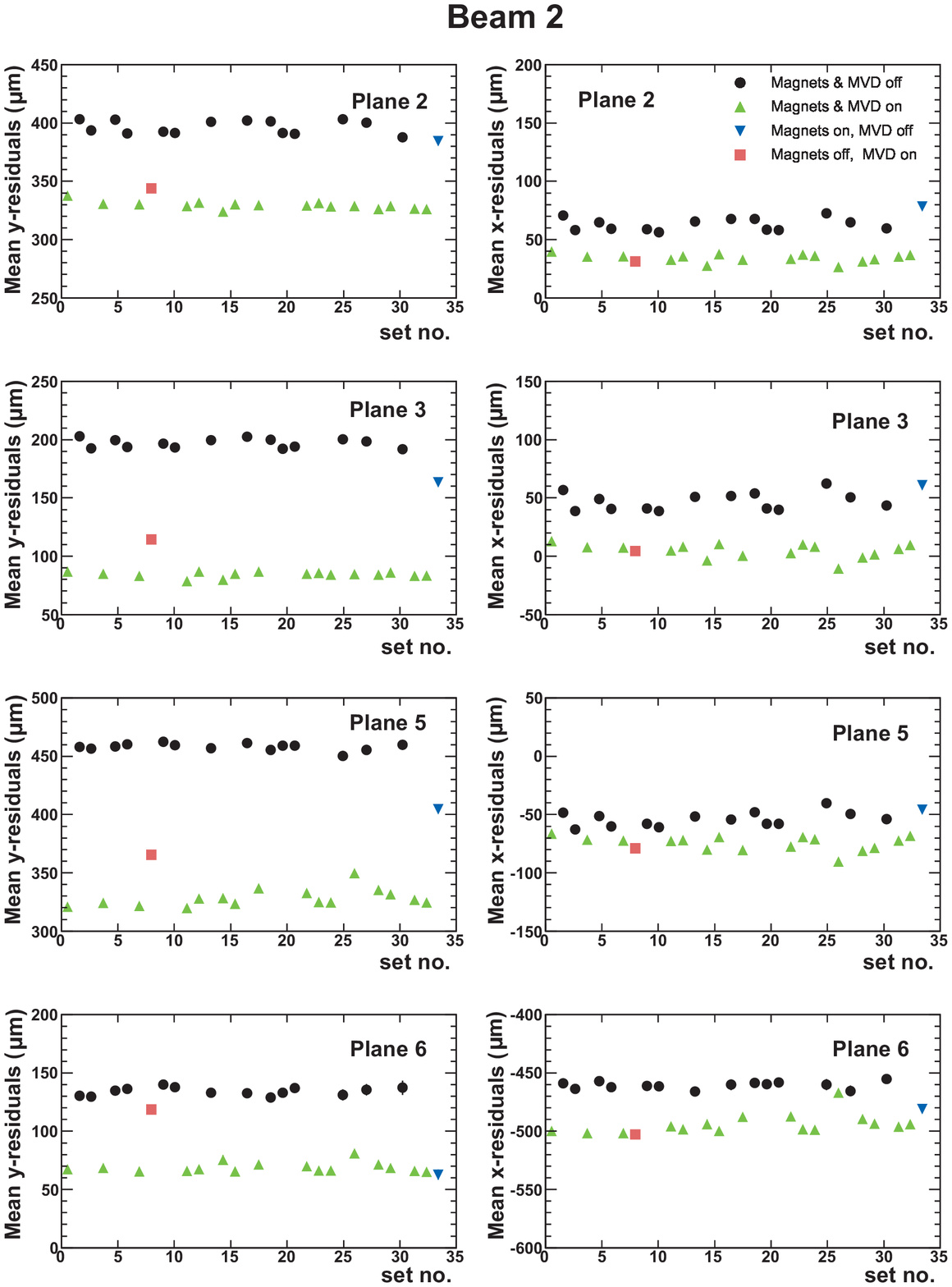}
        \caption{Mean values of ZEUS global-$y$ (left-hand plots) and global-$x$ 
(right-hand plots) residuals for beam-2 during periods of stability. The symbols 
have the same meaning as for Fig.~\ref{fig:mean-res-local}.} 
        \label{fig:mean-res-global}
    \end{center}
\end{figure}

Fig.~\ref{fig:mean-res-global} shows the same results but now plotted for $x$- and 
$y$-residuals in the ZEUS coordinate frame. At this beam position, the size of the 
shift between ON and OFF positions is largest in the $y$ direction but there is also
a smaller but non-zero shift along the $x$ direction. 
Two other positions are shown as single points on these plots: the $8^{\rm th}$ point, 
shown as a square, is for the MVD on and GG magnet off and the last point of all 
(inverted triangle) is for the MVD off and GG magnet on. 

To give a more quantitative estimate of the movements, the beam-2 means from 
Figs~\ref{fig:mean-res-local} and \ref{fig:mean-res-global}
have been averaged over the 16 ON periods and the 14 OFF periods shown. 
The difference in the mean 
values for a given beam and sensor position shows the size of the movement and the 
standard deviations give a good idea of how precisely the MVD support frame is able to 
return to a particular position. These results are collected in 
Table~\ref{tab:beam2-bothxy-on-off-diff}.
Movements of over $100\,\mu$m in the local frame are seen, with the largest standard
deviations on the difference in position around $10\,\mu$m. A similar analysis has been
performed for beam-4 residuals and the results are shown in 
Table~\ref{tab:beam4-bothxy-on-off-diff}. The movements seen at the position of beam-4 are
smaller than those at beam-2, but there is clear evidence for two stable positions with
standard deviations on the differences of position again at most $10\,\mu$m.

\begin{table}[h]
\caption{Mean and standard deviation, in microns, for the average of ON and OFF 
residuals for beam-2 at planes 2, 3, 5 and 6 and their differences in the local 
and global coordinate systems.}
\label{tab:beam2-bothxy-on-off-diff}
\vskip0.3cm
\begin{center}
\begin{tabular}{|c||c|c|c|c||c|c|c|c|}
\hline
LOCAL~~~& P2-$y$ & P3-$y$ & P5-$y$ & P6-$y$ & P2-$x$ & P3-$x$ & P5-$x$ & P6-$x$   \\
\hline
ON& -306.2 & -76.4 & -253.9 & 173.2 & 125.8 & 35.7 & 221.1 & 467.7 \\ 
  &    3.9 & 3.9 & 4.1 & 7.4 & 2.5 & 5.3 & 8.8 & 5.9 \\
\hline
OFF& -379.2 & -195.8 & -378.0 & 99.9 & 132.1 & 51.8 & 264.4 & 469.3 \\
   &  7.5 & 6.8 & 2.2 & 3.4 & 2.5 & 5.4 & 6.8 & 3.1 \\
\hline
DIFF& 73.0 & 119.4 & 124.1 & 73.3 & 6.3 & 16.1 & 43.3 & 1.7 \\
   &  8.4 & 7.9 & 4.7 & 8.2 & 3.5 & 7.5 & 11.1 & 6.6 \\
\hline
\end{tabular}
\end{center}
\vskip0.3cm
\begin{center}
\begin{tabular}{|c||c|c|c|c||c|c|c|c|}
\hline
GLOBAL& P2-$y$ & P3-$y$ & P5-$y$ & P6-$y$ & P2-$x$ & P3-$x$ & P5-$x$ & P6-$x$   \\
\hline
ON& 329.3 & 84.2 & 328.3& 68.8 & 34.1 & 4.7 & -74.6 & -493.9 \\ 
  &  3.1 & 2.4 & 7.5 & 4.3 & 3.5 & 6.1 & 6.2 & 8.5\\
\hline
OFF& 396.7 & 197.0 & 458.1 & 134.1 & 63.0 & 47.1 & -53.9 & -460.5\\
   &  5.8 & 3.9 & 3.0 & 3.4 & 5.3 & 7.6 & 6.2 & 3.1 \\
\hline
DIFF& 67.4 & 112.8 & 129.7 & 65.3 & 28.9 & 42.4 & 20.7 & 33.4 \\
   &  6.6 & 4.6 & 8.1 & 5.4 & 6.3 & 9.8 & 8.8 & 9.0 \\
\hline
\end{tabular}
\end{center}
\end{table}

\begin{table}[h]
\caption{Mean and standard deviation, in microns, for the average of ON and OFF 
residuals for beam-4 at planes 2, 3, 5 and 6 and their differences in the local 
and global coordinate systems.}
\label{tab:beam4-bothxy-on-off-diff}
\vskip0.3cm
\begin{center}
\begin{tabular}{|c||c|c|c|c||c|c|c|c|}
\hline
LOCAL~~~ & P2-$y$ & P3-$y$ & P5-$y$ & P6-$y$ & P2-$x$ & P3-$x$ & P5-$x$ & P6-$x$   \\
\hline
ON& 271.0 & 453.7 & -153.8 & 90.7 & -232.0 & -28.0 & -85.3 & -14.6 \\ 
  &    5.2 & 6.6 & 3.6 & 1.5 & 2.0 & 2.7 & 1.9 & 7.4 \\
\hline
OFF& 240.2 & 421.0 & -185.0 & 107.1 & -246.7 & -54.6 & -109.4 & -46.3 \\
   &  7.0 & 8.8 & 7.5 & 3.0 & 4.4 & 5.4 & 4.7 & 8.3 \\
\hline
DIFF& 30.8 & 32.7 & 31.2 & 16.4 & 14.7 & 26.6 & 24.1 & 31.7 \\
   &  8.7 & 11.0 & 8.3 & 3.4 & 4.9 & 6.1 & 5.0 & 11.1 \\
\hline
\end{tabular}
\end{center}
\vskip0.3cm
\begin{center}
\begin{tabular}{|c||c|c|c|c||c|c|c|c|}
\hline
GLOBAL& P2-$y$ & P3-$y$ & P5-$y$ & P6-$y$ & P2-$x$ & P3-$x$ & P5-$x$ & P6-$x$   \\
\hline
ON& 292.9 & 454.2 & -144.4 & 91.8 & 203.7 & -17.6 & 100.3 & 5.4 \\ 
  & 5.1 & 6.5 & 3.7 & 1.2 & 2.1 & 2.9 & 1.7 & 7.4   \\
\hline
OFF& 263.7 & 424.4 & -173.1 & 111.2 & 221.4 & 12.1 & 127.4 & 35.3 \\
   & 7.2 & 9.1 & 7.7 & 2.8 & 4.1 & 5.0 & 4.2 & 8.4  \\
\hline
DIFF& 29.1 & 29.8 & 28.6 & 19.5 & 17.7 & 29.7 & 27.1 & 29.9 \\
    & 8.8 & 11.2 & 8.6 & 3.0 & 4.6 & 5.8 & 4.6 & 11.2  \\
\hline
\end{tabular}
\end{center}
\end{table}

%\input summary
%%
%% summary
%%

\section{Summary}\label{sec:summary}

This paper has described the laser alignment system for the ZEUS microvertex detector
and given a summary of its performance. The infra-red lasers and semi-transparent
sensors provide five `straightness monitors' that can detect movement of the MVD
support structure at the level of $100\,\mu$m or better. 
The system has been working reliably since it was installed
and commissioned in 2001. The one less than satisfactory feature is the way that
the optical fibre heads are mounted on the central tracking detector. Due to
constraints of time, space and money, a rather crude system had to be employed which
made it difficult to adjust the pointing of the laser beams, even at the time of
installation, and impossible thereafter. 

Time-series analysis of the residuals of the laser beam position with respect to the 
straight lines derived from the positions of a beam at the sensors mounted on the front
and rear of the central tracking chamber is how the laser alignment information is
used.

From the studies reported in this paper, the following conclusions may be drawn:
\begin{itemize}
\item
The MVD support structure is very stable and there is no indication of any large long-term
movement or step change (over the period 2003 to 2007);
\item
When external conditions vary, particularly the GG magnet on or off and the MVD
on-detector electronics being powered or not, the MVD support structure shows
movements locally as large as $100\,\mu$m;
\item
Once the conditions return to the previous state the MVD support tube
returns to its previous configuration, to within $10\,\mu$m. 
\end {itemize}

Finally, it is clear that track data for precision alignment should be collected under 
the operational conditions of regular e-p interaction data taking.

\section*{Acknowledgements}
We thank H.~Kroha and the Munich group for help and advice during the early stages
of this project. We thank T.~Handford for his invaluable help throughout. 
We thank J.~Hill for outstanding effort during the design, construction and installation 
of the laser system. We thank C.~Band, D.~Smith and the Oxford Photo-Fabrication Unit 
for their help with the design and manufacture of the flat readout cables. We thank 
S.~Boogert, and M.~Rigby for their work on the prototype system and L. Hung for her 
help in analysing early production data. Finally we thank C.~Youngman for help and 
advice on how to access HERA and ZEUS slow-control data.
%
%    \input acknowledgements
%\input reference
%%
%% references
%%

% A useful Journal macro
% Author, Journal name, volume, year, page
\def\Journal#1#2#3#4{{#1} {\bf #2}, (#3) #4}

% Some useful journal names
\def\NCA{\em Nuovo Cimento}
\def\NIM{\em Nucl. Instrum. Methods}
\def\NIMA{{\em Nucl. Instrum. Methods} A}
\def\NPB{{\em Nucl. Phys.} B}
\def\PLB{{\em Phys. Lett.} B}
\def\PRL{\em Phys. Rev. Lett.}
\def\PRD{{\em Phys. Rev.} D}
\def\ZPC{{\em Z. Phys.} C}
\def\EPC{{\em Eur. Phys. J.} C}
\def\IEEE{{\em IEEE Trans. Nucl. Sci.}}

\end{document}